# Forecasting Megaelectron-Volt Electrons inside Earth's Outer Radiation Belt: PreMevE 2.0 Based on Supervised Machine Learning Algorithms

**Rafael Pires de Lima[1,2], Yue Chen[1], and Youzuo Lin[1]**

[1]Los Alamos National Laboratory, Los Alamos, NM, USA,

[2]The University of Oklahoma, School of Geosciences, OK, USA

Corresponding authors: Rafael Pires de Lima (rlima@ou.edu) and Yue Chen (cheny@lanl.gov)

**Key Points:**

- Several linear and artificial neural network models are tested for forecasting MeV electron events inside Earth's radiation belt

- New PreMevE 2.0 model makes 1- and 2-day forecasts of 1 MeV electron events with high fidelity

- Relationship between trapped 1 MeV electrons and precipitating electrons appears to be dominated by linear components

**Abstract**

Here we present the recent progress in upgrading a predictive model for Megaelectron-Volt (MeV) electrons inside the Earth's outer Van Allen belt. This updated model, called PreMevE 2.0, is demonstrated to make much improved forecasts, particularly at outer Lshells, by including upstream solar wind speeds to the model's input parameter list. Furthermore, based on several kinds of linear and artificial machine learning algorithms, a list of models were constructed, trained, validated and tested with 42-month MeV electron observations from Van Allen Probes. Out-of-sample test results from these models show that, with optimized model hyperparameters and input parameter combinations, the top performer from each category of models has the similar capability of making reliable 1-day (2-day) forecasts with Lshell-averaged performance efficiency values ∼ 0.87 (∼ 0.82). Interestingly, the linear regression model is often the most successful one when compared to other models, which suggests the relationship between dynamics of trapped 1 MeV electrons and precipitating electrons is dominated by linear components. It is also shown that PreMevE 2.0 can reasonably well predict the onsets of MeV electron events in 2-day forecasts. This improved PreMevE model is driven by observations from longstanding space infrastructure (including a NOAA satellite in low-Earth-orbit, the solar wind monitor at the L1 point, and one LANL satellite in geosynchronous orbit) to make high-fidelity forecasts for MeV electrons, and thus can be an invaluable space weather forecasting tool for the future.

## 1. Introduction

Man-made satellites operating in medium- and high-altitude Earth orbits are continuously exposed to hazardous space radiation originated from different sources. Among them, one major contributor is the relativistic electron population—with energies comparable to and/or larger than their rest energy of 0.511 Megaelectron-volt (MeV)—trapped inside Earth's outer Van Allen belt. Owning to their high penetration capability, these MeV electrons are difficult to be fully stopped by normal shielding. Particularly, during MeV electron events when electron intensities across the outer belt are greatly enhanced to sustaining high levels, space-borne electronic systems with inadequate hardening are susceptible to deep-dielectric charging and discharging phenomenon caused by those electrons (Lai et al., 2018), and thus may suffer severe damages or even stop functioning. Therefore, protecting critical space infrastructures from harsh space weather conditions– including MeV electron events – has high priority for stakeholders such as the space industry, service providers and government agencies.

Similar to terrestrial weather services, real-time monitoring and model forecasting are the two principle ways of mitigating risks from outer-belt MeV electrons. Given the successful NASA Van Allen Probes mission, previously known as RBSP (Mauk et al., 2013), quickly approaches its end, the need of reliable forecasting models for MeV electrons becomes compelling once again due to the coming absence of in-situ measurements. Indeed, forecasting models have been developed including such as SPACECAST framework (Horne et al., 2013) for the whole outer radiation belt, and Relativistic Electron Forecast Model (based on Baker et al. (1990)) currently operated by NOAA specifically for electrons at geosynchronous (GEO) orbit. Recently, Chen et al. (2019) has developed and verified a new predictive MeV electron model called PreMevE to forecast MeV electron events throughout the whole outer radiation belt, using simple linear filters with inputs mainly from low-Earth-orbit (LEO) observations. In this work, we further improve PreMevE model by applying and testing several supervised machine learning algorithms with optimized selection of input parameters.

Machine learning (ML) has been a topic in consideration for more than half a century (e.g., Minsky, 1961; Hartigan & Wong, 1979; Hopfield, 1982), and its popularity increased significantly in the last decade with numerous applications in various research fields. Examples of success include seismicity studies (e.g., Kortström et al., 2016; Perol et al., 2018; Sinha et al., 2018; Ren et al., 2019; Wang et al., 2019), geological mapping (e.g. Cracknell & Reading, 2014; Pires de Lima & Marfurt, 2018), optical/electrical geoscientific images classification (e.g. Duarte-Coronado et al., 2019; Pires de Lima et al., 2019a; Pires de Lima et al., 2019b; Valentín et al., 2019), medical image segmentation and classification (e.g., Ronneberger et al., 2015; Tajbakhsh et al., 2016; Qayyum et al., 2017), speech recognition (e.g., Graves & Schmidhuber, 2005; Graves et al., 2013), and etc. Among them, as observed by LeCun et al. (2015), the work of Krizhevsky et al. (2012) was the breakthrough responsive for the rapid adoption of deep learning by the computer vision and others communities.

Meanwhile, the application of ML also has gained momentum in the space weather community. An early use of artificial neural networks to predict the flux of energetic electrons at GEO orbit was presented by Stringer et al. (1996) in which GOES-7 data were used to make one-hour nowcasts of hourly-averaged fluxes of electrons at energies of 3-5 MeV. Later, Ukhorskiy et al. (2004) and Kitamura et al. (2011) used artificial neural networks to develop one-day forecasts of daily averaged electron fluxes at GEO. More recently, Shin et al. (2016) used a neural network scheme with solar wind inputs to predict GEO electrons over a wide energy range with different time resolutions. Wei et al. (2018) also successfully improved the one-day forecasts of >2 MeV electron fluxes at GEO by applying deep learning algorithms. For a review, Camporeale (2019) has summarized the recent progresses and opportunities of applying ML for space weather forecasting problems, including predicting geomagnetic indices, relativistic electrons, solar flares occurrence, coronal mass ejection propagation time, solar wind speed and etc.

The purpose of this work is to present how PreMevE has been upgraded with ML algorithms to make improved predictions of MeV electron flux distributions. With no requirement of in-situ MeV electron measurements except for at GEO, this unique model has shown its great potential of meeting the predictive requirements for outer-belt electrons during the post-RBSP era. Section 2 briefly describes data and parameters to be used for this study, and the selected ML algorithms and their implementations are explained in Section 3. Section 4 compares and summarizes the prediction performance of different models, followed by detailed discussions in Section 5. This work is concluded by Section 6 with a summary of our findings and possible future directions.

## 2. Data and Input Parameters

Electron data used in this work include observations made by particle instruments aboard a RBSP spacecraft, one Los Alamos National Laboratory (LANL) GEO satellite, and one NOAA Polar Operational Environmental Satellite (POES) in a time period ranging from February 2013 to August 2016, as shown in Figure 1. Electron data used here are the same as in Chen et al. (2019) in which detailed descriptions of the original data and their preparation can be found, and here is a brief recap. First, trapped 1 MeV electrons across a range of L-shells (L ≤ 6) are in situ measured by the Magnetic Electron Ion Spectrometer (MagEIS) instrument (Blake et al., 2013) on board RBSP-a, and the spin-averaged fluxes are plotted in Panel A as a function of Lshell and time. Here we use McIlwain's L values (McIlwain, 1966) calculated from the quiet Olson and Pfitzer magnetic field model (Olson & Pfitzer, 1977) together with the International Geomagnetic Reference Field model. At GEO, we use observations from the Synchronous Orbit Particle Analyzer (SOPA, Belian et al., 1992) instrument carried by the GEO satellite LANL-01A. For simplicity, all GEO fluxes are put on the fixed L = 6.6 and plotted in the top of Panel A. Then, precipitating electrons are monitored by the Space Environment Monitor 2 (SEM2) instruments on board NOAA POES satellites in low-Earth-orbits (LEOs, Evans et al., 2000), and the count rates from the 90° telescopes on NOAA-15 are presented for three energy channels as in Panels B, C and D. Here L values for NOAA-15 are calculated from the International Geomagnetic Reference Field model. Additionally, upstream solar wind (SW) speeds in Panel E are downloaded from CDAweb site and added to models' inputs. All RBSP-a, LANL-01A, and POES-15 electron fluxes as well as solar wind speeds in Figure 1 are binned by 5 hours to allow for RBSP's full coverage on the outer belt for each time bin. The Lshell bin size for electrons is 0.1.

Throughout this work, we refer to POES electron fluxes at > 100 keV, > 300 keV, and > 1000 keV as E2, E3, and P6 respectively. Logarithmic values of E2, E3, and P6, along with standardized scaled values of SW speeds form the input data sets, or the predictors, being used to forecast the logarithm of 1 MeV trapped electron fluxes, sometimes also referred to as "target". The standardized of SW is done by subtracting the mean and dividing by the standard deviation (both the mean of 404.8 km/s and standard variation of 86.8 km/s are computed with the training set as defined in Section 4). Hereinafter, when we refer to 1 MeV target, E2, E3, and P6 fluxes, we are actually referring to their logarithmic values. Lshell coverage of this study is confined to 2.8 – 6 (the range of RBSP) and 6.6 (LANL GEO), while fluxes at other Lshells can be derived by radial interpolation or extrapolation (Chen et al., 2019).

## 3. Supervised Machine Learning Algorithms

ML can be described as a collection of techniques in which systems improve their performance through automatic analysis of data. The power of ML models lies in their capacity to extract statistical information (patterns and features) from ample data with no requirement of hypothesis, in a sharp contrast to physics-based models in which researchers manually select parameters to be used as input for models with specific governing physics. ML models are capable of extracting signature and correspondence that might be overlooked by traditional methods, e.g., nonlinear relationship, and can be relatively easy to use with multiple input sources. Therefore, under certain circumstances, ML models can outperform traditional ones. For example, Tajbakhsh et al. (2016) found that deep neural network models outperformed handcrafted solutions in medical image analysis tasks. Nevertheless, one major drawback of ML models, particularly deep neural networks, is its incomplete capability in interpretability ("how") and explainability ("why") (Murdoch et al., 2019). Thus, sometimes ML models can be complicated to explain, hindering our ability to propose new theories based on ML results.

Common ML algorithm types include supervised, unsupervised, semi-supervised, and reinforcement learning (Ayodele, 2010). Algorithms used here fall under the category of supervised learning as they make use of input sample data paired with an appropriate label. The label here refers to 1 MeV electron flux at different Lshells, the target value to be forecasted. Moreover, the models implemented here can be classified as regressions, as the labels are specified scalar values.

As explained by Camporeale (2019), supervised regressors try to find the mapping relationship between a set of multidimensional inputs $\boldsymbol{x} = (x_1, x_2, \ldots, x_N)$ and its corresponding scalar output label $y$, under the general form

$$y = f(\boldsymbol{x}) + \epsilon, \tag{1}$$

where $f: \mathbb{R}^N \to \mathbb{R}$ is a linear or nonlinear function and $\epsilon$ represents additive noise. All methods used to find the unknown function $f$ can be seen as an optimization problem where the objective is to minimize a given loss function. The loss function is a function that maps the distance between all the predicted and target values into a real number, therefore providing some "cost" associated with the prediction. The following four subsections provide details in each one of the supervised regressor models used in this study. A comprehensive discussion on artificial neural networks and deep learning models can be found in LeCun et al. (2015), with information about techniques common to several artificial intelligence applications. To exemplify the supervised learning problem as a flux forecasting task, consider predicting the 1 MeV electron fluxes at time $t$ at GEO shell using the past values of 1 MeV electron fluxes at GEO. Suppose we use $M$ training samples to perform the analysis, and the number of past values we wish to use for each time step is four ($N = 4$). That is, we have $M$ pairs of $(\boldsymbol{x}_t, y_t)$ training samples, or $\{(\boldsymbol{x}_1, y_1), (\boldsymbol{x}_2, y_2) \ldots, (\boldsymbol{x}_M, y_M)\}$ where $\boldsymbol{x}_t = (x_{t-1}, x_{t-2}, x_{t-3}, x_{t-4})^T \in \mathbb{R}^{N=4}$ and $y_t \in \mathbb{R}$. We can rewrite the predictors $\boldsymbol{x}_t$ as a matrix $X \in \mathbb{R}^{N \times M}$, where each column of the matrix $X$ represents one $\boldsymbol{x}_t$ training sample vector. The $y_t$ samples can also be defined as a single row matrix $Y \in \mathbb{R}^{1 \times M}$. The goal of ML training is to optimize the internal parameter values of the given mapping function $f$—a specified ML algorithm—by minimizing the loss function associated with the noise matrix $\epsilon$ after inserting $X$ and $Y$ back into Eq. (1). Here we use the past values of multiple input data, including E2, E3, P6, and solar wind speed, to forecast 1 MeV electron fluxes at each individual L-shell. Next, we describe the four selected algorithms including linear regression, multilayer perceptron, convolutional neural network, and long short-term memory methods.

## 3.1 Linear Regression

Linear regression is the simplest supervised learning method, while sometimes it is also interpreted as the simplest ML algorithm. This algorithm has a vast range of applications and constitutes a basic building block for more complex algorithms. The linear regression equation is given by

$$f(\boldsymbol{x}_i) = \boldsymbol{w}^T \boldsymbol{x}_i + \boldsymbol{b}, \tag{2}$$

where $\boldsymbol{w}$ is a vector containing weights and $\boldsymbol{b}$ is the bias term. In a predictive problem, $y$ as in Eq. (1) represents the label, or target, to be predicted (the 1 MeV electron flux), $\boldsymbol{x}$ represents the input data (e.g. past values of precipitating electron fluxes) and $\boldsymbol{w}$ represents the set of linear coefficients that minimize the loss, or the sum of the errors between all true values of $y$ and the predicted $f(\boldsymbol{x}_i)$. From the optimization perspective, the weights $\boldsymbol{w}$ can be obtained using a simple ordinary least squares method. Linear models are simple models generally very useful as baselines, and their selection for this work is also due to the success of previous work by Chen et al. (2019).

## 3.2 Multilayer Perceptron

Starting from the linear model, a single neuron can be defined as

$$f = a(\boldsymbol{w}^T \boldsymbol{x_i} + \boldsymbol{b}), \qquad (3)$$

where $a(.)$ is an element-wise activation function. The activation function is responsible to introduce non-linearity to the model. Some of the most common activation functions are the Rectified Linear Unit (ReLU, Hahnloser et al., 2000; Nair & Hinton, 2010) and the Exponential Linear Unit (ELU, Clevert et al., 2015). ReLU is a piecewise linear function that outputs the input for positive values, zero otherwise; ELU outputs the identity for positive values as well, however ELU uses a logarithm curve for negative values ($constant(exp(input) - \boldsymbol{1})$). A hidden layer is a set of neurons, or units, that take in a set of inputs ($\boldsymbol{x}$) and produce an output $f$. If we use the $f_i^{[l]}$ notation to represent the output of the neuron $i$ at layer $l$, we can write Eq. (3) for the following layer as $f^{l+1} = a(\boldsymbol{w}^T \boldsymbol{f}^{[l]} + \boldsymbol{b})$ to represent the inputs for layer $l$ +1 depend on the output of layer $l$. Figure 2 illustrates a single neuron in the left and how sets of neurons can be combined to form layers and neural networks in the right. Here the information flows from left (the input) to right (the output). This structure is a class of Feedforward Networks, sometimes named multilayer perceptron (MLP). Loosely defined, an artificial neural network (NN) is a model consisting of connected neurons. The term deep model or deep learning is generally used for NNs containing more than one hidden layers. When all neurons in a layer receive input from all elements in the previous layer (e.g. the hidden layers in Figure2b), they are also called fully or densely connected layers.

### 3.3 Convolutional Neural Networks

Convolutional neural networks (CNNs) are powerful and influential deep learning model architectures. The computer vision field strongly adopted CNNs as their workforce after the CNN described in Krizhevsky et al. (2012) has achieved new levels of accuracy in the popular ImageNet Large Scale Visual Recognition Competition (Russakovsky et al., 2015). All CNNs make use of the fundamental convolutional kernel. Convolution operates on two functions, one generally interpreted as the "input", and the other as a "filter". The filter is commonly referred to as "kernel". The kernel is applied on the input, producing an output image or signal. During the training stage, the values of kernels are updated in such a way that the output generated by the CNN is more similar to the desired label, i.e., minimizes the cost. Just like the neurons described in subsection 3.2, a set of convolutional kernels can be combined into layers. Dumoulin & Visin (2016) showed details on the arithmetic of convolutions for deep learning. Here, we provide only the essential equation for 1D convolution. A 1D convolution of the input vector $\boldsymbol{x}$ and the kernel $\boldsymbol{g}$ of length $m$ is given by

$$(x * g)(i) = \sum_{j=1}^{m} g(j) x \left( i - j + \frac{m}{2} \right). \qquad (4)$$

A CNN unit in deep learning models is a composite of activation function and the convolution term in Eq. (4), i.e., $f(i) = a((x * g)(i))$.

Springenberg et al. (2014) observed that CNNs commonly use alternating convolution and max-pooling layers followed by a small number of fully connected layers. The models are typically regularized during training by using dropout. Max-pooling are simple down-sampling steps in which the maximum value for each patch (containing multiple values) of a feature is used to represent the entire patch, effectively reducing the feature size. Dropouts layers randomly select a percentage of their inputs to be ignored during the training phase. Dropouts are useful to avoid overfitting. Dropout is a general approach and not specific for CNN models. Srivastava et al. (2014) showed that dropout improves the performance of NNs on many supervised learning tasks such as speech recognition, document classification, vision and computational biology.

### 3.4 Long Short-term Memory

Long short-term memory (LSTM) networks are a popular recurrent neural network (RNN) structure introduced by Hochreiter & Schmidhuber (1997). RNN is a class of artificial neural networks in which neurons can be connected to form a directed graph along a temporal sequence (Figure 3). Different from traditional feedforward NNs, LSTM has internal loops to allow to retain information from previous time steps and decide its usage for predictions. Indeed, the LSTM basic unit is called memory cell inside which internal components can decide when to keep or override information in the memory cell, when to access the information in memory cell, and when to prevent other units from being perturbated (Hochreiter & Schmidhuber, 1997). Olah (2015) provides a detailed walkthrough of the LSTM components. LSTMs are constantly used in speech recognition problems (e.g. Graves et al., 2013; Graves & Schmidhuber, 2005) as well as forecasting (e.g. Kong et al., 2019). Here LSTM was selected for testing as a representative of RNNs.

### 4. Testing Algorithms and Model Performance

Following ML best practices, we split the data into training, validation, and test sets. The training set is the data effectively used for model optimization. The validation set is used to tune model hyperparameters, such as the number of neurons/layers or optimization options. Finally, the test set is reserved for model performance evaluation on the final stage. Here, the training data set consists of observations in the first 4,008 time bins (roughly 835 days, or 27.4 months, 65% of the whole data set), the validation set has observations for the next 841 time bins (roughly 175 days, or 5.8 months, 14% of the data), and the test set is for the final 1,280 time bins (roughly 267 days, or 8.8 months, 21% of the data). Observational data are split in such a manner so that the major observational gap over days 840 – 850 is in between the sets, thus the models are always trained, validated, and tested in segments containing continuous observations.

The optimization goal for all the models is to reduce the root-mean-square error (RMSE) between the real value $y$ and the predicted value $f$, both with the size $M$. RMSE is defined as $\sqrt{\frac{\Sigma_{j=1}^{M}(f_j-y_j)^2}{M}}$. In this study, linear models minimize the error using the ordinary least squares, while artificial NN models use Adam optimization as defined by Kingma & Ba (2014).

Chen et al. (2019) has demonstrated that E2 fluxes can be used for predicting the onset timings of MeV electron events, and here we also computed the normalized temporal derivatives of E2 fluxes, naming it dE2, and tested by adding it to the input data sets for predicting onsets. The dE2 at time bin $t$ for E2 is defined as $dE2_t = \frac{E2_t - E2_{t-1}}{E2_{t-1}}$. The temporal correlation between E2, dE2, and trapped MeV electron fluxes can be recognized from Figure 4.

### 4.1 Test Input Parameter Combinations

Our first experiment tests different combinations of input parameters with the objective to find the set of input data that can best predict 1 MeV electrons. Specifically, we use Linear and LSTM models to evaluate what combination of input parameters yields the highest Performance Efficiency (PE). PE provides a measure of quantifying the accuracy of predictions by comparing to variance of the target. Naming $y$ as the true value (the logarithm of the target 1 MeV electron flux) and $f$ as the predicted value, both with size $M$, PE is defined as

$$PE = 1 - \frac{\sum_{j=1}^{M}(y_j - f_j)^2}{\sum_{j=1}^{M}(y_j - \bar{y})^2},$$ (5)

where $\bar{y}$ is the mean of $\boldsymbol{y}$. PE does not have a lower bound, and the perfect score is 1.0, meaning all predicted value perfectly match observed data, or that $\boldsymbol{f = y}$.

To make 1-day (25 hr) forecasts of MeV electrons for a single Lshell, our models ingest the past values of the input data at the same L-shell. The only exception is at GEO, where the model inputs also include past MeV electron fluxes at GEO from in-situ measurements. Additionally, as Chen et al. (2019) found E2, E3, and P6 values at GEO have relatively weak correlations with 1 MeV electrons, E2, E3, and P6 channels at L-shell of 4.6 are used instead for model inputs. The term "window size" refers to how many five-hour time bins of input data are needed by the models. Chen et al. (2019) found a window size of 15 time bins (equivalent to 75 hours) to be effective for the forecast of MeV electrons. Adhering to the "power of two" ML convention, here we used a window size of 16. The "power of two" rule is based on the fact that CPUs and GPUs memory architecture are usually organized in powers of two, thus using power of two data organization can be beneficial for computation efficiency. For naming convention, when a LSTM model has one layer with 128 memory cells, we use LSTM-128 as the name for this model; the linear models are referred to as LinearReg throughout the manuscript. Here results from the submodel 1 and 2 of previous PreMevE in Chen et al. (2019) are always cited as linear1 and linear2 for a baseline comparison. Note in this work all PE values and predicted fluxes from linear1 and linear2 are for 1-day forecasts only.

Table 1 summarizes the overall PE values (averaged over all Lshells) for twenty tests performed for 1-day predictions. For each of the two categories of models, ten input parameter sets are tested, starting from each single parameter to various combinations. Here we focus on the out-of-sample PE values, i.e., those in the column of PE val+test, to judge model performance. The general trend is that more parameters lead to better performance. For example, the last LinearReg model with all parameters as input (the 10[th] model) has not only the highest overall PE value of 0.861 but also the highest PE at GEO (0.587). These two values are higher than those for linear2 (0.797 and 0.352), which indicates significant improvements. (Linear1 was designed for capturing onset timings of MeV electron events and thus its PE values are always lower than those of linear2 (Chen et al., 2019).) Interestingly, the last two LSTM models (19[th] and 20[th]) have the highest overall and GEO PE values for this category, but still slightly lower than those of the 10[th] LinearReg model.

In this step, we also confirmed that adding SW speeds to the input list improves model performance, which was not tested previously in Chen et al. (2019). In Table 1, the overall PE for the 1[st] model by using SW speed as the sole input parameter is 0.518, which suggests this simple model can predict MeV electrons over the whole outer belt to some degree but not as well as the linear2, although the PE of 0.557 at GEO is much higher than that of linear2 (0.352). In comparison, PE values from the 11[th] model show that using SW speed as the sole parameter for LSTM model does not work as good as for the 1[st] LinearReg model particularly at GEO. When comparing models without and with SW speeds, e.g., the 2[nd] vs 7[th] (12[th] vs 17[th]) and 8[th] vs 9[th] (18[th] vs 19[th]), improvements in overall PE are 0.009 (0.017) and 0.005 (0.013), respectively, while the improvements in PE at GEO are more significant up to 0.11. We also tested the dE2 and its addition to the input has effects less significant than SW speeds when comparing the PE values of the 10[th] (20[th]) model to those of the 9[th] (19[th]).

Details of how model PEs improve as a function of Lshell are presented in Figure 5. For models in both categories, the top performer has much higher PE values than those of linear2 across the whole belt, with the most significant improvements for outer Lshells $> \sim 4.5$ and the maximum increment of $> 0.4$ at $L \sim 5.5$. It can be clearly seen from the green curve in Panel A that the SW speed is a very helpful predictor at outer Lshells ($L > \sim 5$) especially for LinearReg models, but inefficient for inner Lshells. This can be explained by the fact that in the high Lshell region particle dynamics are more controlled by adiabatic effects, and is also consistent with the experience from existing predictive models for electrons at GEO (e.g., Baker et al., 1990). However, as in Panel B, the LSTM model using SW speeds as the sole predictor has only a few L-shells with PE values greater than zero. In summary, results in both Table 1 and Figure 5 suggest that the model PE values are higher with the use of more input data from multiple precipitating electron channels as well as the SW speed. Therefore, tests in the rest of this study generally use the parameter combination including all inputs.

## 4.2 Model Selection and Evaluation Metrics

We then advanced to test a list of models built upon different algorithms with varying model hyperparameters (e.g., the window size and number of neurons). There are four different categories of models—Linear, MLP, LSTM, and CNNs—as described in Section 3, and here are how these models and test runs were set up. First, to account for cross-shell information as in Chen et al. (2019), some tests include E2 data at the Lshell of 4.6 as input for all other L-shells. Second, all MLP models presented here are composed of two hidden layers—the first one has 64 neurons and the second has 32 neurons, and the neurons use ELU as the activation function. In our early testing, we discovered ELU achieving marginally better performance than the most adopted ReLU activation function. A dropout layer that randomly selects 50% of the input to be inactivated after each one of the activation functions is included to help prevent overfitting. The output layer consists of a single neuron without an activation function. The dropout layer, used during training and deactivated during prediction, and the output layer are not accounted as hidden layers, but are also part of the model. We name such models MLP-64-32-elu. Then, CNN models with a window size 16 are composed of two convolutional layers, the first convolutional layer contains 64 kernels followed by a Max-pooling layer with size and stride equal two, and the second convolutional layer contains 32 kernels followed by a Max-pooling layer with the same size and stride. The CNN models with a window size 4 are composed of a single convolutional layer with 64 kernels followed by a Max-pooling layer. The kernels are one-dimensional with a size of three and use ReLU as activation function. The convolutional layers are finalized with 50% dropout. The output layer consists of a single neuron without an activation function. Those CNN models are named Conv-64-32 and Conv-64, respectively. Finally, LSTM models follow the same structure as the ones described in Section 4.1.

Model performance is again evaluated by PE values by comparing forecasts to the target data. Table 2 presents the overall PE values for 24 test runs performed for 1-day predictions, and Table 3 presents PE values for the same test runs for 2-days predictions. Inside each category, the effects of window size, neuron/layer numbers and input parameters are tested and compared, and Table 1 and 2 only show results of models with good performance. For 1-day forecasts as in Table 2, the 6th LinearReg model has the high overall PE of 0.872 for out-of-sample test and 0.587 at GEO. Top performers in the other three categories have similar overall and GEO PE values. All those values are higher than the overall PE of 0.797 and GEO PE of 0.352 from linear2 for 1-day forecasts. For 2-day forecasts in Table 3, top performers are the same as for 1-day predictions except for the MLP category. Here, the 6th LinearReg model has the highest overall PE of 0.827 for out-of-sample test and 0.333 at GEO. Again, top performers have overall PEs ~0.82 for 2-day predictions, which is lower than the ~0.87 for 1-day predictions but still higher than the ~0.80 of linear2 for 1-day forecasts. (Chen et al. (2019) has shown that the linear2 have lower PE for 2-day forecasts than 1-day.) Their PEs at GEO are mostly above 0.33, comparable to linear2. Note for the MLP category, the 9th model is the top performer in Table 3, with no E2 at L=4.6 for input—instead of the 11th in Table 2. None CNN models in Table 3 can make 2-day forecasts at GEO very well.

Figure 6 plots PE curves for both 1- and 2-day forecasts as a function of Lshell, which further confirm our models' performance are more robust than previous results published in Chen et al. (2019). First, the PE curves for all four top models cluster together, with PE values at outer L-shells (minimum > ~0.3) lower than those at inner L-shells (maximum > 0.8 in left panel and >0.7 for right). All PE curves for both 1-day and 2-day are well above that from linear2 (1-day) expect at low Lshells for 2-day forecasts. The most significant improvements in PE are for Lshells > 4.5. For 1-day forecasts, due to the addition of E2 at L=4.6 to the parameter list, the 6th LinearReg model in Table 2 (the green thick line in Figure 6A) can be seen to outperform with higher overall PE than the 10th model in Table 1. In addition, the performance of LinearReg models is persistently good for both 1- and 2-day forecasts, particularly at GEO where other top models degrades quickly as in Figure 6.

It is striking how the models (LinearReg, MLP, LSTM, CNN) show very similar forecasting ability when using similar input data. Plus, the LinearReg models seem to have leading performance for the forecasting in many scenarios, particularly for 2-day predictions. Two main observations should be taken for such behaviors. The first one is that a great part of the interplay between trapped 1 MeV electrons and input parameters (precipitating electrons and SW speeds) appear to be mostly linear. Previous PreMevE in Chen et al. (2019) has high PE using linear filters to forecast MeV electrons, and our findings corroborate previous results. The second observation is that artificial NNs, as depicted in Section 3, have their linear component. As a linear model achieves good results, artificial neural networks are expected to do at least the same. Thus, the dominance of linear components explains why the top models from all four categories of algorithms have very similar predictive performance. In addition, the secondary role from non-linear components makes CNN models having the best overall PE of 0.877 for validation and test set combined as in Table 2, as well as the MLP and LSTM models having the best PE at GEO (Tables 2 and 3). Therefore, this new PreMevE 2.0 model indeed includes all four algorithms, which form an ensemble of predictive models whose relative weights are left to future work for determination. Next, we take a closer look at predicted results from all four algorithms.

## 5 Detailed Predictions and Discussions

An overview of the 1-day forecasted flux distributions is exhibited in Figure 7 compared to the 1 MeV flux target. Visually, forecasted distributions from the four top performers as in Table 2 (Panels B-D) resemble the observations (Panel A) very closely. Portions of data used for training, validation, and test are marked out by color bars in the bottom of the figure. It can be seen that the enhancement of fluxes, elevated flux levels (red regions), and decay afterwards during each individual MeV electron event are reproduced very well, although the dropouts of MeV electrons (blue strips) at large Lshells are not well captured sometimes (e.g., the one on ~ day 1080) or even totally missed (e.g., the one on ~ day 870). It is deemed acceptable at this stage since PreMevE model mainly aims to forecast high flux levels of MeV electrons. Similarly, Figure 8 compares 2-day forecasted results to target data and shows an akin resemblance, confirming the stable predictive performance of PreMevE 2.0 with a longer lead time.

Furthermore, Figure 9 shows even more details how closely the 1-day forecasted fluxes are compared to the targets over the combined validation and test period for selected L-shells. Here flux curves from the same models as in Figure 7 as well as linear2 model are plotted. The four PreMevE 2.0 model curves pack together tightly and trace the target curve (black) closely, particularly during decays of high intensity events. The closeness between the target and each forecasted curve depicts the performance of each model. A close inspection reveals that the linear2 curve (yellow) is often the one farthest away from the target, showing as almost the envelop line of the predictions, while the LinearReg curve (green) appears the closest tracer of target at L=4.5, and the MLP curve (red) is the winner for other two Lshells. Nevertheless, it can be seen that the forecasted values often lag behind the target at onsets of MeV electron events, e.g., the ones on ~ day 988 and 1093 at L = 4.5.

Figure 10 illustrates how well the onsets of MeV electron events at L=4.5 are captured by the models. Here forecasts from linear1 is also plotted in blue for comparison. (Linear1, or the submodel 1, in Chen et al. (2019) is specifically designed to predict the onsets.) We selected 16 major events in which MeV electron fluxes increase by > ~10 times, marked out by the vertical gray boxes in Figure 10. Linear1 (the blue curve) successfully predicts the onsets of all major MeV electron events at this Lshell, indicated by the leading edges of significant sudden increments in fluxes fallen within the boxes with a width of 25 hr (also called prediction windows). In comparison, although the four models (particularly the LinearReg model in green) often predict onsets earlier than linear2, they only successfully predict eight of them (those marked with green letter Y), fail seven, and have one event barely making inside the prediction widow. In other words, 1-day forecasts from PreMevE 2.0 predict the onsets at L =4.5 with a success rate below 50%, which is better than linear2 but far behind linear1.

Two-day forecasts are also presented in Figures 11 and 12. Again, forecasted results at three Lshells in Figure 11 closely trace the target, similar to Figure 9. Interestingly, for all 16 selected major MeV electron events in Figure 12, the onsets of 11 events are successfully predicted by the four models at L = 4.5, while the failed events decrease to 4. This increases the success rate of onset prediction to ~70%. Judged from this number and above results, this new PreMevE 2.0 model is able to combine the advantages of both linear1 and linear2 by not only predicting the arrivals of new MeV electrons but also specifying evolving flux levels closely, which is an encouraging progress.

Results from LinearReg and LSTM models at GEO are specifically presented in Figure 13 for both 1- and 2-day predictions. For 1-day forecasts in the top three panels, it can be seen that fluxes from LinearReg (green) and LSTM (purple) trace observations (black) more closely than linear2 (yellow), consistent with their higher PE values as shown in Table 2. Also, forecasts from LinpearReg and LSTM appear to predict the onsets of MeV electron events at about the same level as linear1, by comparing the leading edges of flux spikes in those curves. For 2-day forecasts, LinearReg PE value in Table 3 suggests that 2-day forecasts from LinearReg model are close to 1-day forecasts from linear2, which can be seen from the entangled LinearReg and linear2 curves as in Panels D-F. Forecasts from the LSTM model is not as good, although they still capture the general trend of 1 MeV electrons at GEO.

Despite Chen et al.'s (2019) and our time window selection, the question of for how long history of each particle population can significantly affect MeV electron prediction remains open. Figure 14 shows the Spearman correlation of the input data (E2, E3, P6, and SW speed) and the target (1 MeV electrons) for three selected L-shells using different time lags. Spearman correlation does not assume that the data follows a particular distribution, so it is a non-parametric measure of monotonic relationship. The results in Figure 14 show that the Spearman correlation between input and target decays with longer time lags. The correlation remains stronger for longer periods at inner L-shells (i.e., longer memory) and decays faster for outer L-shells (shorter memory). We also note the correlation between SW speeds and target gets more significant when moving to outer L-shells, which is consistent from our discussions in Section 4.1. Curiously, the shape the correlation curve of E3 is similar to the shape of P6, whereas the shape of E2 is similar to the shape of SW. All these suggest more robust models can be elaborated with a variation of inputs (window sizes and parameter combinations) for different Lshells. In fact, Figure 5 shows that E2+SW are apparently the best combination of input for outer shell prediction, whereas other combination of input presents stronger values of PE for inner shells. Given a threshold value of ~0.4 for significant correlation, it is seen from Figure 14 that a fixed window size for all Lshells may range from > ~14 (to include the maximum correlation values) up to ~20 (to avoid too long history).

Previously Chen et al. (2019) used 300 time bins to train linear1 and linear2 submodels to forecast MeV electrons. Table 1 shows that linear1 and linear2 models have a weaker forecasting performance than the LinearReg models trained with the similar inputs. The difference in performance can be explained by the fact that a much larger training set incorporate a wider flux variations that can be helpful to train the models. Besides, the addition of SW speeds definitely helps improve the performance of linear models at large Lshells.

In this work, we have performed tests on the number of units, types of activation functions, and number of layers, though it is still possible that a more intensive artificial NN architecture testing will find a more appropriate model for MeV electron forecasting. Moreover, we expect that more available data can be useful to improve models' performance. We plan to test with observations over longer period as well as for higher energy electrons in the next step.

## 6 Summary and Conclusions

This new PreMevE 2.0 model aims to forecast MeV electron distributions even with no in-situ measurements available, e.g., during the post-RBSP era, and it is designed to be driven by easily accessible inputs from long-standing satellite constellations in LEO and GEO as well as at the Lagrangian 1 point of Sun-Earth system. Meanwhile, deep learning algorithms have recently achieved new state-of-the-art accuracy in many problems partially due to the increase of observations. Therefore, it is reasonable for us to foresee an increase in both performance and use of deep learning model for MeV electron forecasting as more space weather data have been accumulated and made available.

In this work, we have tested (1) different model input parameter combinations and (2) four categories of supervised machine learning algorithms, with the goal of upgrading our predictive model for MeV electrons inside the Earth's outer radiation belt. This new PreMevE 2.0 model has been demonstrated to make much improved forecasts, particularly at large Lshells, by including upstream solar wind speeds to the model's inputs. Additionally, based on four categories of linear and artificial machine learning algorithms, a list of models were constructed, trained, validated and tested with 42-month MeV electron observations from NASA Van Allen Probes mission. Model predictions over the 14-month long out-of-sample test show that, with optimized model hyperparameters and input parameter combinations, the top performer from each category of models has the similar capability of making reliable 1- and 2-day forecasts with Lshell-averaged performance efficiency values of ~ 0.87 and ~0.82, respectively. Interestingly, the linear regression model is often the most successful one when compared to other models, which suggests the relationship between the dynamics of trapped 1 MeV electrons and precipitating electrons is dominated by linear components. It is also shown that PreMevE 2.0 can predict the onsets of MeV electron events in 2-day forecasts with a reasonable success rate of ~70%. This improved PreMevE model is driven by observations from existing space infrastructure (including a NOAA LEO satellite, the solar wind monitor at L1 point, and one LANL GEO satellite) to make high-fidelity forecasts for MeV electrons, and thus can be an invaluable space weather forecasting tool for the community.

## Acknowledgments, Samples, and Data

The authors declare no conflicts of interest. Pires de Lima acknowledges CNPq (grant no. 203589/2014-9) for graduate sponsorship. We gratefully acknowledge the support of NASA Heliophysics Space Weather Operations to Research Program (18-HSWO2R18-0006), the NASA Heliophysics Guest Investigators program (14-GIVABR14_2-0028), and LANL internal funding. We want to acknowledge the PIs and instrument teams of NOAA POES SEM2 and RBSP MagEIS for providing measurements and allowing us to use their data. Thanks to CDAWeb for providing OMNI data. RBSP and POES data used in this work were downloadable from the missions' public data websites (https://www.rbsp-ect.lanl.gov and http://www.ngdc.noaa.gov), while LANL GEO SOPA data have been included as supplementary material of Chen et al. (2019) and downloadable from the AGU website.

# References

Ayodele, T. O. (2010). Types of machine learning algorithms. In *New advances in machine learning*. IntechOpen

Baker, D. N., McPherron, R. L., Cayton, T. E., & Klebesadel, R. W. (1990). Linear prediction filter analysis of relativistic electron properties at 6.6 RE. *Journal of Geophysical Research: Space Physics*, *95*(A9), 15133–15140. https://doi.org/10.1029/JA095iA09p15133

Belian, R. D., Gisler, G. R., Cayton, T., & Christensen, R. (1992). High- *Z* energetic particles at geosynchronous orbit during the Great Solar Proton Event Series of October 1989. *Journal of Geophysical Research*, *97*(A11), 16897. https://doi.org/10.1029/92JA01139

Blake, J. B., Carranza, P. A., Claudepierre, S. G., Clemmons, J. H., Crain, W. R., Dotan, Y., et al. (2013). The Magnetic Electron Ion Spectrometer (MagEIS) Instruments Aboard the Radiation Belt Storm Probes (RBSP) Spacecraft. *Space Science Reviews*, *179*(1–4), 383–421. https://doi.org/10.1007/s11214-013-9991-8

Camporeale, E. (2019). The Challenge of Machine Learning in Space Weather: Nowcasting and Forecasting. *Space Weather*, 2018SW002061. https://doi.org/10.1029/2018SW002061

Chen, Y., Reeves, G. D., Fu, X., & Henderson, M. (2019). PreMevE: New Predictive Model for Megaelectron-Volt Electrons Inside Earth's Outer Radiation Belt. *Space Weather*, *17*(3), 438–454. https://doi.org/10.1029/2018SW002095

Clevert, D.-A., Unterthiner, T., & Hochreiter, S. (2015). Fast and Accurate Deep Network Learning by Exponential Linear Units (ELUs). *ArXiv E-Prints*, arXiv:1511.07289.

Cracknell, M. J., & Reading, A. M. (2014). Geological mapping using remote sensing data: A comparison of five machine learning algorithms, their response to variations in the spatial distribution of training data and the use of explicit spatial information. *Computers & Geosciences*, *63*, 22–33. https://doi.org/10.1016/J.CAGEO.2013.10.008

Duarte-Coronado, D., Tellez-Rodriguez, J., Pires de Lima, R., Marfurt, K., & Slatt, R. (2019). Deep convolutional neural networks as an estimator of porosity in thin-section images for unconventional reservoirs. In *SEG Technical Program Expanded Abstracts 2019* (pp. 3181–3184). Society of Exploration Geophysicists. https://doi.org/10.1190/segam2019-3216898.1

Dumoulin, V., & Visin, F. (2016). A guide to convolution arithmetic for deep learning. *ArXiv E-Prints*.

Evans, D. S., Greer, M. S., & (U.S.), S. E. C. (2000). *Polar orbiting environmental satellite space environment monitor-2: instrument description and archive data documentation*. Boulder, CO: U.S. Dept. of Commerce, National Oceanic and Atmospheric Administration, Oceanic and Atmospheric Research Laboratories, Space Environment Center.

Graves, A., & Schmidhuber, J. (2005). Framewise phoneme classification with bidirectional LSTM and other neural network architectures. *Neural Networks*, *18*(5–6), 602–610. https://doi.org/10.1016/J.NEUNET.2005.06.042

Graves, A., Jaitly, N., & Mohamed, A. (2013). Hybrid speech recognition with Deep Bidirectional LSTM. In *2013 IEEE Workshop on Automatic Speech Recognition and Understanding* (pp. 273–278). IEEE. https://doi.org/10.1109/ASRU.2013.6707742

Hahnloser, R. H. R., Sarpeshkar, R., Mahowald, M. A., Douglas, R. J., & Seung, H. S. (2000). Digital selection and analogue amplification coexist in a cortex-inspired silicon circuit. *Nature*, *405*(6789), 947–951. https://doi.org/10.1038/35016072

Hartigan, J. A., & Wong, M. A. (1979). Algorithm AS 136: A K-Means Clustering Algorithm. *Journal of the Royal Statistical Society. Series C (Applied Statistics)*, *28*, 100–108. https://doi.org/10.2307/2346830

Hochreiter, S., & Schmidhuber, J. (1997). Long Short-Term Memory. *Neural Comput.*, *9*(8), 1735–1780. https://doi.org/10.1162/neco.1997.9.8.1735

Hopfield, J. J. (1982). Neural networks and physical systems with emergent collective computational abilities. *Proceedings of the National Academy of Sciences of the United States of America*, *79*(8), 2554–8. https://doi.org/10.1073/pnas.79.8.2554

Horne, R. B., Glauert, S. A., Meredith, N. P., Koskinen, H., Vainio, R., Afanasiev, A., et al. (2013). Forecasting the Earth's radiation belts and modelling solar energetic particle events: Recent results from SPACECAST. *J. Space Weather Space Clim.*, *3*, A20. https://doi.org/10.1051/swsc/2013042

Kingma, D. P., & Ba, J. (2014). Adam: A Method for Stochastic Optimization. *ArXiv E-Prints*, arXiv:1412.6980.

Kitamura, K., Nakamura, Y., Tokumitsu, M., Ishida, Y., & Watari, S. (2011). Prediction of the electron flux environment in geosynchronous orbit using a neural network technique. *Artificial Life and Robotics*, *16*(3), 389–392. https://doi.org/10.1007/s10015-011-0957-1

Kong, W., Dong, Z. Y., Jia, Y., Hill, D. J., Xu, Y., & Zhang, Y. (2019). Short-Term Residential Load Forecasting Based on LSTM Recurrent Neural Network. *IEEE Transactions on Smart Grid*, *10*(1), 841–851. https://doi.org/10.1109/TSG.2017.2753802

Kortström, J., Uski, M., & Tiira, T. (2016). Automatic classification of seismic events within a regional seismograph network. *Computers & Geosciences*, *87*, 22–30. https://doi.org/10.1016/J.CAGEO.2015.11.006

Krizhevsky, A., Sutskever, I., & Hinton, G. E. (2012). ImageNet Classification with Deep Convolutional Neural Networks. In *Proceedings of the 25th International Conference on Neural Information Processing Systems - Volume 1* (pp. 1097–1105). USA: Curran Associates Inc. Retrieved from http://dl.acm.org/citation.cfm?id=2999134.2999257

Lai, S. T., Cahoy, K., Lohmeyer, W., Carlton, A., Aniceto, R., & Minow, J. (2018). Deep Dielectric Charging and Spacecraft Anomalies. In N. Buzulukova (Ed.), *Extreme Events in Geospace* (pp. 419–432). Elsevier. https://doi.org/https://doi.org/10.1016/B978-0-12-812700-1.00016-9

LeCun, Y., Bengio, Y., & Hinton, G. (2015). Deep learning. *Nature*, *521*(7553), 436–444. https://doi.org/10.1038/nature14539

Mauk, B. H., Fox, N. J., Kanekal, S. G., Kessel, R. L., Sibeck, D. G., & Ukhorskiy, A. (2013). Science Objectives and Rationale for the Radiation Belt Storm Probes Mission. *Space Science Reviews*, *179*(1), 3–27. https://doi.org/10.1007/s11214-012-9908-y

McIlwain, C. E. (1966). Magnetic coordinates. *Space Science Reviews*, *5*(5), 585–598. https://doi.org/10.1007/BF00167327

Minsky, M. (1961). Steps toward Artificial Intelligence. *Proceedings of the IRE*, *49*(1), 8–30. https://doi.org/10.1109/JRPROC.1961.287775

Murdoch, W. J., Singh, C., Kumbier, K., Abbasi-Asl, R., & Yu, B. (2019). Definitions, methods, and applications in interpretable machine learning. *Proceedings of the National Academy of Sciences*, *116*(44), 22071 LP – 22080. https://doi.org/10.1073/pnas.1900654116

Nair, V., & Hinton, G. E. (2010). Rectified Linear Units Improve Restricted Boltzmann Machines. In *Proceedings of the 27th International Conference on International Conference on Machine Learning* (pp. 807–814). USA: Omnipress. Retrieved from http://dl.acm.org/citation.cfm?id=3104322.3104425

Olah, C. (2015). Understanding LSTM Networks. Retrieved from http://colah.github.io/posts/2015-08-Understanding-LSTMs/

Olson, W. P., & Pfitzer, K. A. (1977). *Magnetospheric magnetic field modeling. Annual scientific report*. Huntington Beach.

Perol, T., Gharbi, M., & Denolle, M. (2018). Convolutional neural network for earthquake detection and location. *Science Advances*, *4*(2), e1700578. https://doi.org/10.1126/sciadv.1700578

Pires de Lima, R., & Marfurt, K. J. (2018). Principal component analysis and K-means analysis of airborne gamma-ray spectrometry surveys. In *SEG Technical Program Expanded Abstracts 2018* (pp. 2277–2281). Society of Exploration Geophysicists. https://doi.org/10.1190/segam2018-2996506.1

Pires de Lima, R., Suriamin, F., Marfurt, K. J., & Pranter, M. J. (2019). Convolutional neural networks as aid in core lithofacies classification. *Interpretation*, *7*(3), SF27–SF40. https://doi.org/10.1190/INT-2018-0245.1

Pires de Lima, R., Bonar, A., Coronado, D. D., Marfurt, K., & Nicholson, C. (2019). Deep convolutional neural networks as a geological image classification tool. *The Sedimentary Record*, *17*(2), 4–9. https://doi.org/10.210/sedred.2019.2

Qayyum, A., Anwar, S. M., Awais, M., & Majid, M. (2017). Medical image retrieval using deep convolutional neural network. *Neurocomputing*, *266*, 8–20. https://doi.org/10.1016/J.NEUCOM.2017.05.025

Ren, C. X., Dorostkar, O., Rouet-Leduc, B., Hulbert, C., Strebel, D., Guyer, R. A., et al. (2019). Machine Learning Reveals the State of Intermittent Frictional Dynamics in a Sheared Granular Fault. *Geophysical Research Letters*, *46*(13), 7395–7403. https://doi.org/10.1029/2019GL082706

Ronneberger, O., Fischer, P., & Brox, T. (2015). U-Net: Convolutional Networks for Biomedical Image Segmentation. In N. Navab, J. Hornegger, W. M. Wells, & A. F. Frangi (Eds.), *Medical Image Computing and Computer-Assisted Intervention -- MICCAI 2015* (pp. 234–241). Cham: Springer International Publishing.

Russakovsky, O., Deng, J., Su, H., Krause, J., Satheesh, S., Ma, S., et al. (2015). ImageNet Large Scale Visual Recognition Challenge. *International Journal of Computer Vision*, *115*(3), 211–252. https://doi.org/10.1007/s11263-015-0816-y

Shin, D.-K., Lee, D.-Y., Kim, K.-C., Hwang, J., & Kim, J. (2016). Artificial neural network

prediction model for geosynchronous electron fluxes: Dependence on satellite position and particle energy. *Space Weather*, *14*(4), 313–321. https://doi.org/10.1002/2015SW001359

Sinha, S., Wen, Y., Pires de Lima, R. A., & Marfurt, K. (2018). Statistical controls on induced seismicity. Unconventional Resources Technology Conference. https://doi.org/10.15530/urtec-2018-2897507-MS

Springenberg, J. T., Dosovitskiy, A., Brox, T., & Riedmiller, M. (2014). Striving for Simplicity: The All Convolutional Net. *ArXiv E-Prints*, arXiv:1412.6806.

Srivastava, N., Hinton, G., Krizhevsky, A., Sutskever, I., & Salakhutdinov, R. (2014). Dropout: A Simple Way to Prevent Neural Networks from Overfitting. *Journal of Machine Learning Research*, *15*(1), 1929–1958. Retrieved from http://dl.acm.org/citation.cfm?id=2627435.2670313

Stringer, G. A., Heuten, I., Salazar, C., & Stokes, B. (1996). Artificial Neural Network (ANN) Forecasting of Energetic Electrons at Geosynchronous Orbit (pp. 291–295). American Geophysical Union (AGU). https://doi.org/10.1029/GM097p0291

Tajbakhsh, N., Shin, J. Y., Gurudu, S. R., Hurst, R. T., Kendall, C. B., Gotway, M. B., & Liang, J. (2016). Convolutional Neural Networks for Medical Image Analysis: Full Training or Fine Tuning? *IEEE Transactions on Medical Imaging*, *35*(5), 1299–1312. https://doi.org/10.1109/TMI.2016.2535302

Ukhorskiy, A. Y., Sitnov, M. I., Sharma, A. S., Anderson, B. J., Ohtani, S., & Lui, A. T. Y. (2004). Data-derived forecasting model for relativistic electron intensity at geosynchronous orbit. *Geophysical Research Letters*, *31*(9), L09806. https://doi.org/10.1029/2004GL019616

Valentín, M. B., Bom, C. R., Coelho, J. M., Correia, M. D., de Albuquerque, M. P., de Albuquerque, M. P., & Faria, E. L. (2019). A deep residual convolutional neural network for automatic lithological facies identification of Brazilian pre-salt oilfield wellbore image logs. *Journal of Petroleum Science and Engineering*. https://doi.org/10.1016/J.PETROL.2019.04.030

Wang, T., Zhang, Z., & Li, Y. (2019). EarthquakeGen: Earthquake generator using generative adversarial networks. In *SEG Technical Program Expanded Abstracts 2019* (pp. 2674–2678). Society of Exploration Geophysicists. https://doi.org/10.1190/segam2019-3216687.1

Wei, L., Zhong, Q., Lin, R., Wang, J., Liu, S., & Cao, Y. (2018). Quantitative Prediction of High-Energy Electron Integral Flux at Geostationary Orbit Based on Deep Learning. *Space Weather*, *16*(7), 903–916. https://doi.org/10.1029/2018SW001829

**Figures**

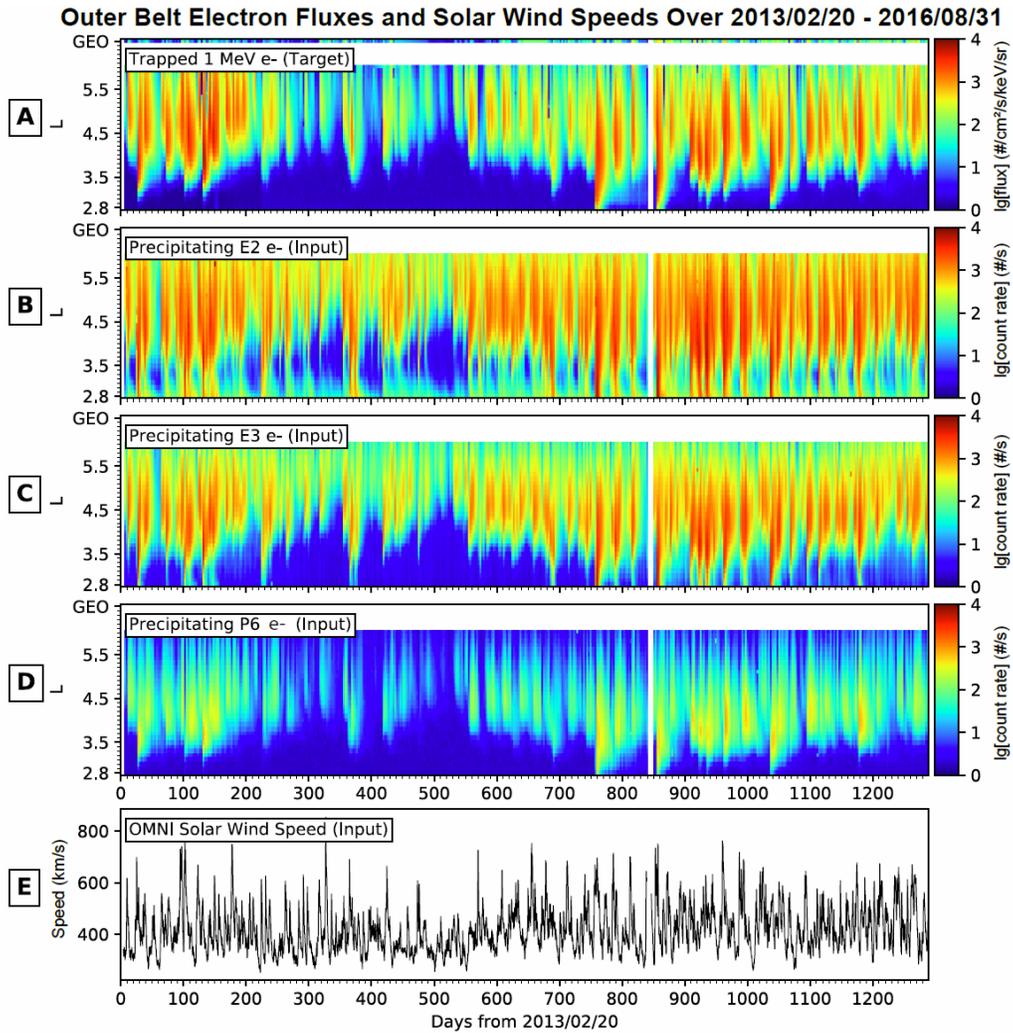

**Figure 1**: **Overview of electron observations and solar wind speeds used in this study.** All panels present for the same 1289-day interval starting from 2013/02/20. Panel **A** shows flux distributions of 1 MeV electrons, the variable to be forecasted (i.e., targets). Similarly, **B**, **C**, and **D** show count rates of precipitating electrons measured by NOAA-15 in a low-Earth-orbit, for E2, E3, and P6 channels respectively. **E** plots the solar wind speeds measured upstream of the magnetosphere as in the OMNI data set for the period. Data in Panels B-E are model inputs (i.e., predictors).

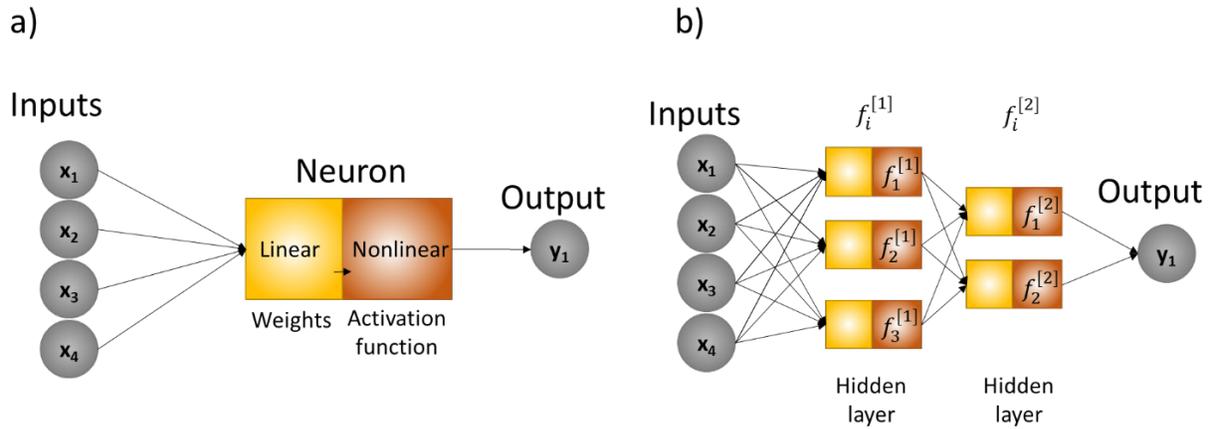

**Figure 2: Visual generic representation of a single neuron and an artificial neural network. a)** shows a single neuron that can be split into linear and nonlinear components, as well as the input and output data. In the case of a forecasting problem, the inputs can be data representing past times $t_{-1}, t_{-2}, t_{-3}, t_{-4}$, and the output is prediction at current time $t_0$ or even some future time. **b)** shows how a set of neurons constitute a layer and how the output of a layer can be used as input for the next layer.

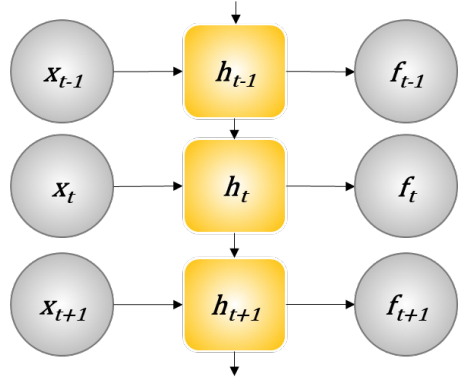

**Figure 3: Representation of a recurrent neural network.** In LSTM models, the basic unit $h$ is also called a memory cell. The input vector $x$ at an arbitrary time $t$ is processed by a memory cell $h$ which produces an output $f(x)$. The output produced by $h_{t-1}$ is also part of the input for $h_t$. Thus, events at time $t$ are processed with information from the previous steps. The output produced by $h$ can be used as input to the next layer just like the described for the previous models.

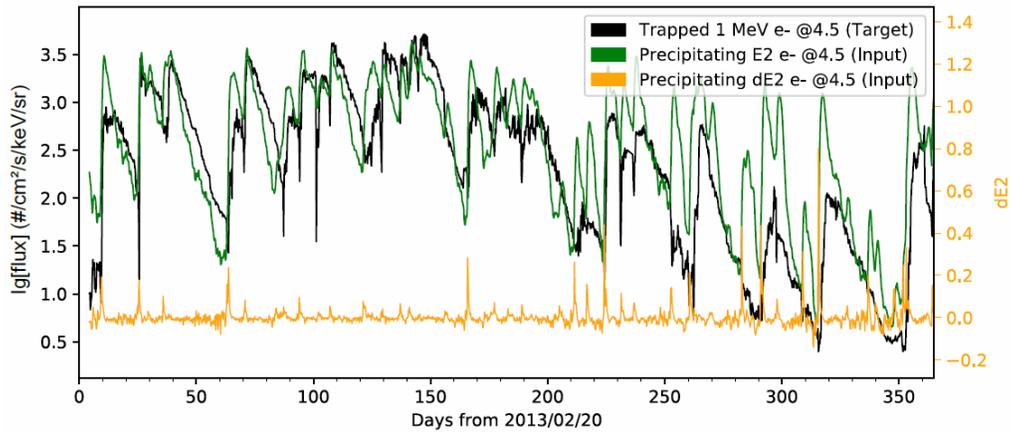

**Figure 4. Temporal correlation between E2, dE2, and 1 MeV electrons fluxes in the first year of the interval.** Note the leading edges of E2 increments (green) and the spikes in dE2 (yellow) generally precede the onsets of MeV electron events with a significant one-to-one temporal relationship.

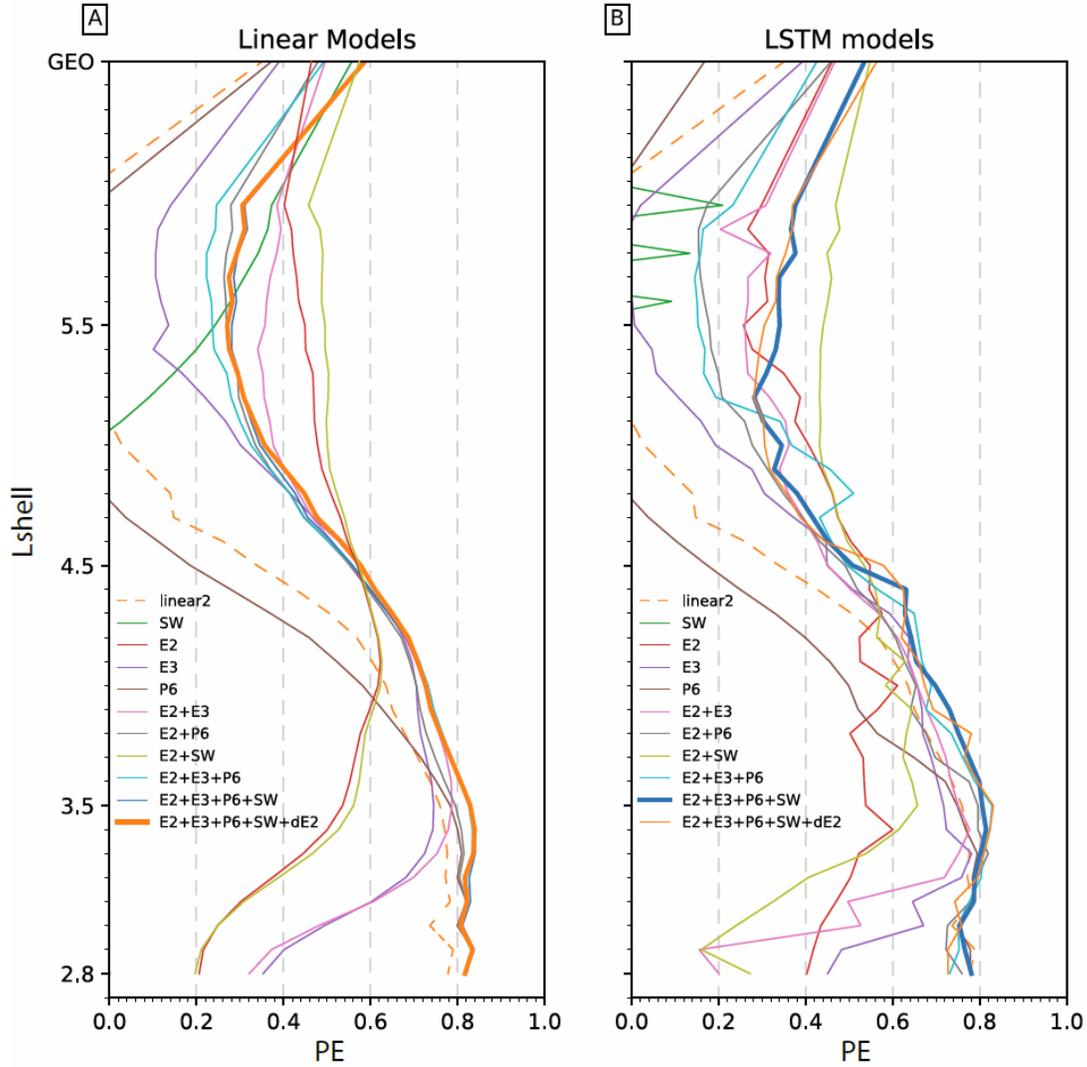

**Figure 5. PE values for the combined validation and test sets are presented as a function of Lshell for different models and input parameters as in Table 1. A**. LinearReg models and **B**. LSTM models. PE curve for linear2 model (dashed) is plotted for comparison. The model with the best performance—highest overall PE—for each category is highlighted with a thick line.

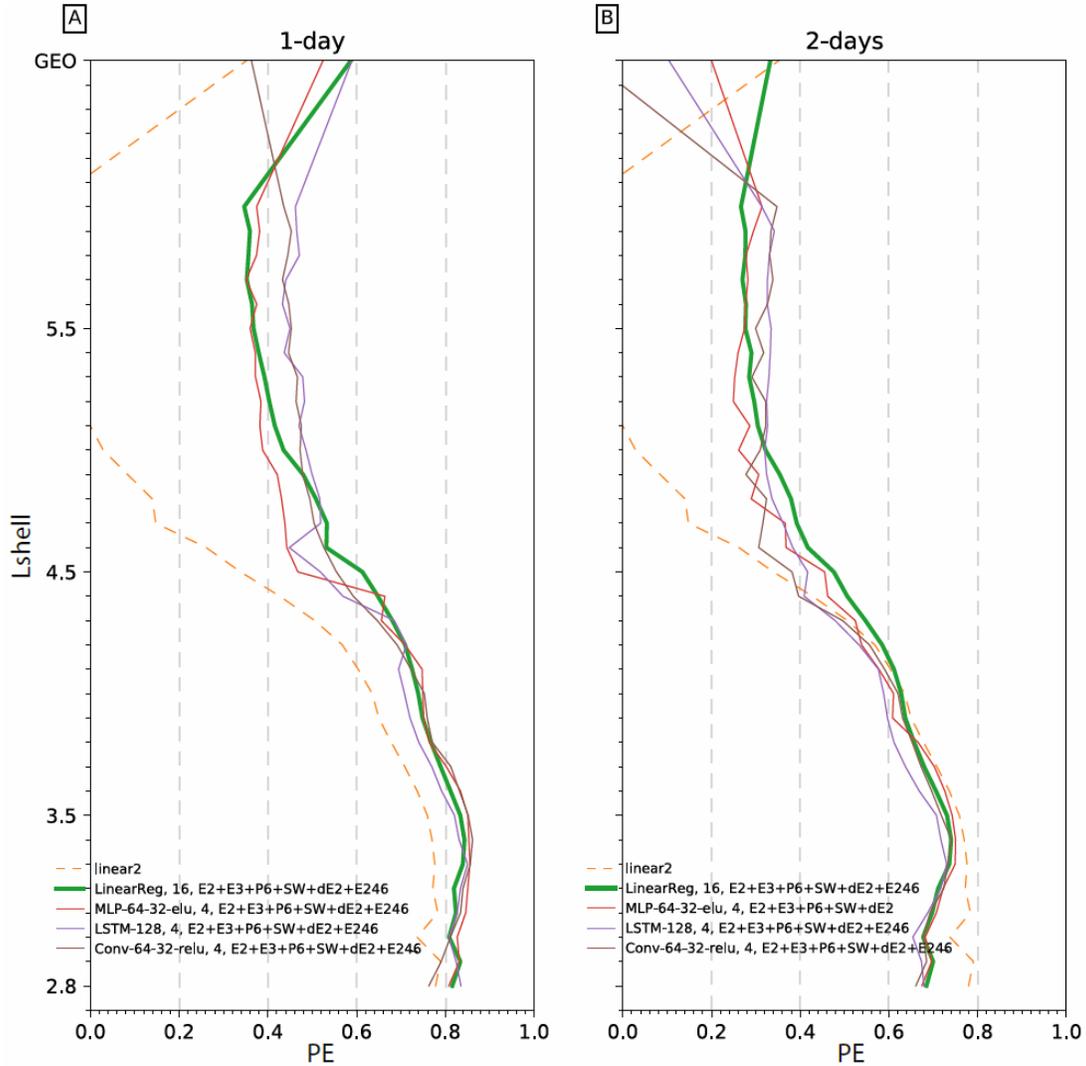

**Figure 6. Model PE values over the combined validation and test data sets are presented as a function of Lshell for the top performers in Table 2 and 3. A.** Top performer of each category for 1-day forecasts. PE curve of linear 2 for 1-day forecasts is plotted in dashed line for comparison. **B.** Top performer of each category for 2-days forecasts. Note the dashed line is still linear2 for 1-day forecasts. LinearReg models are highlighted in thick lines in both panels.

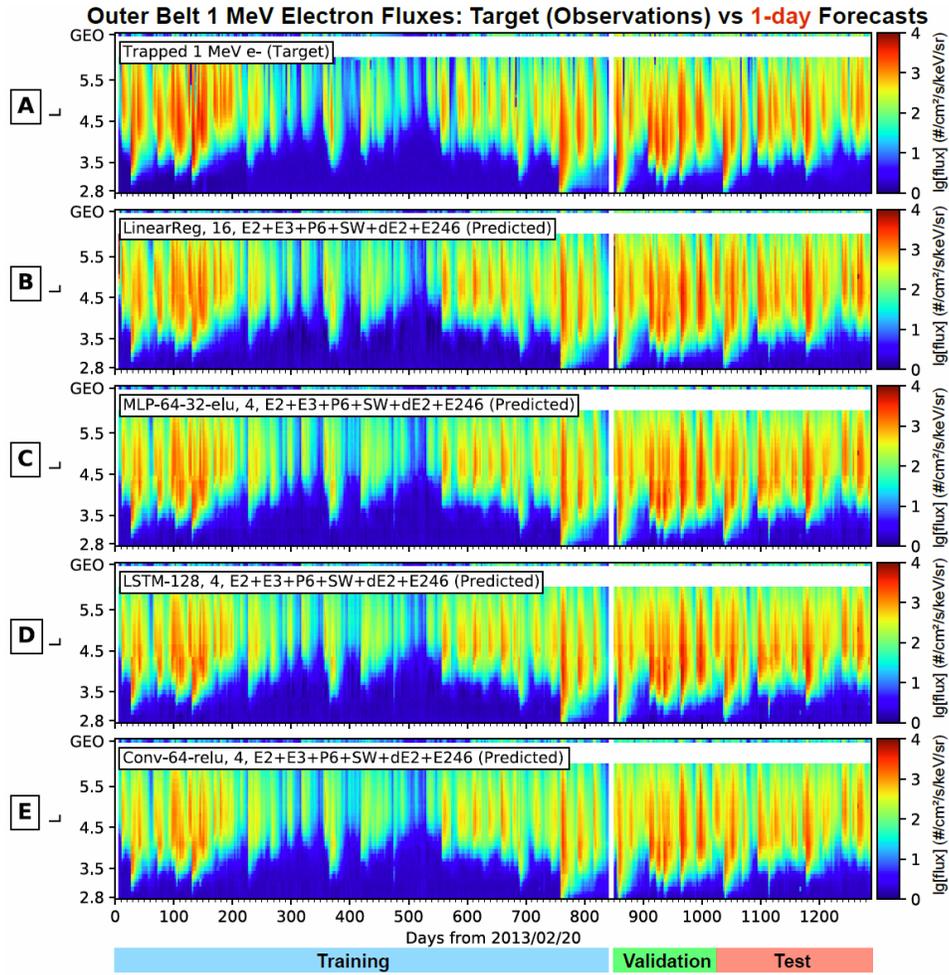

**Figure 7. Overview of target and 1-day forecasted fluxes across all Lshells. A** shows the observed flux distributions to be forecasted for 1 MeV electrons. **B, C, D,** and **E** show, respectively, predictions from the models with the highest overall PE including linear regression model, MLP, LSTM, and CNN models.

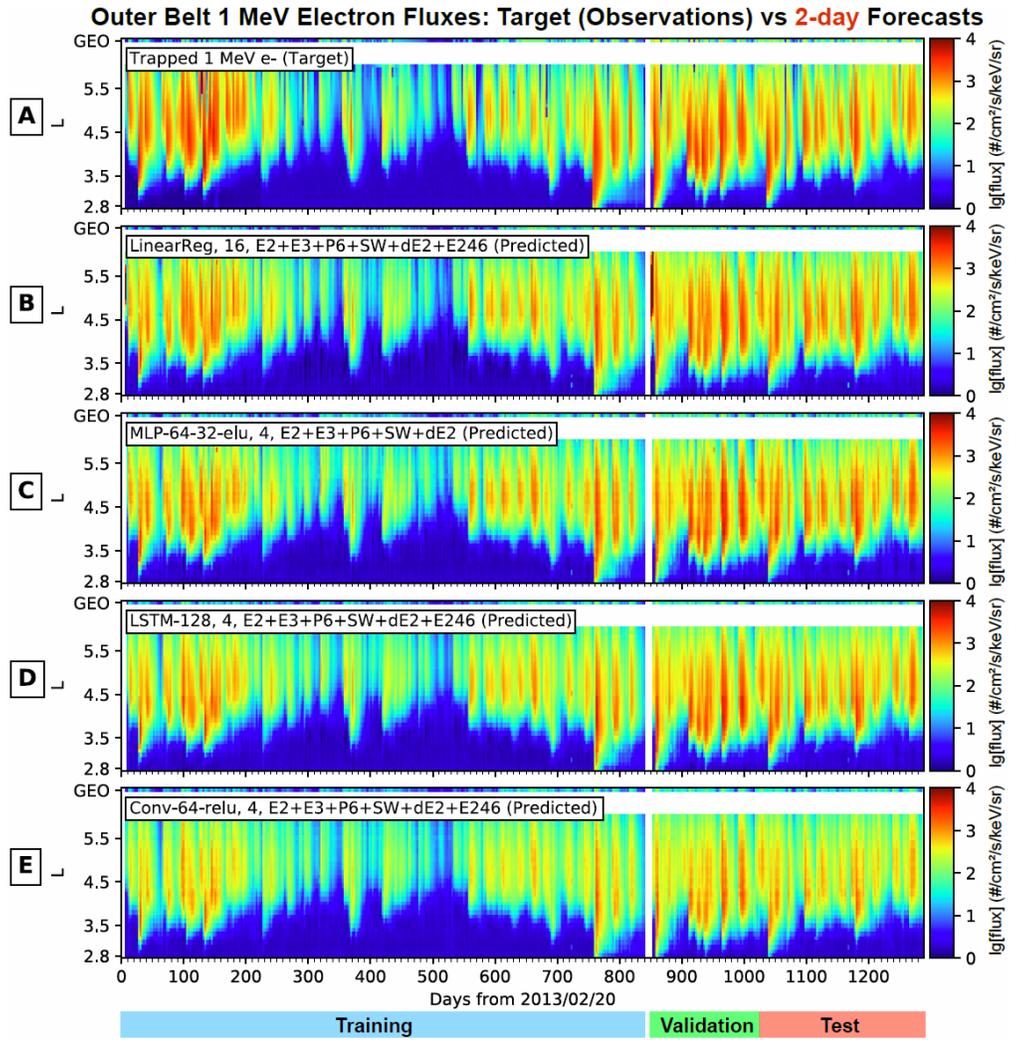

**Figure 8. Overview of target and 2-day forecasted fluxes across all Lshells.** All panels are in the same format as in Figure 7.

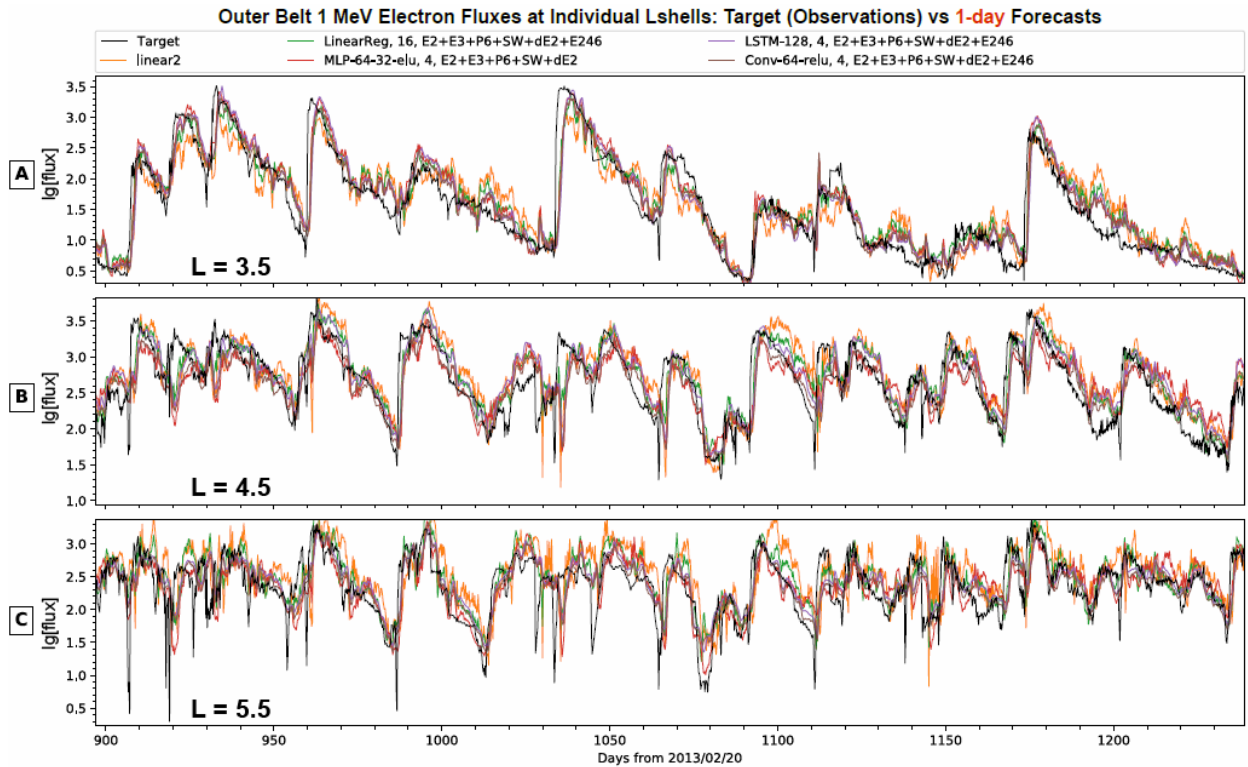

**Figure 9. One-day forecasts compared to target fluxes at three selected Lshells over the combined validation and test period.** Panels **A, B,** and **C** are for Lshells of 3.5, 4.5, and 5.5, respectively. The measured 1 MeV electrons (black) are compared to predictions from the LinearReg, MLP, LSTM, and CNN models with highest PE in each category (Table 1) as well as linear2 model (yellow).

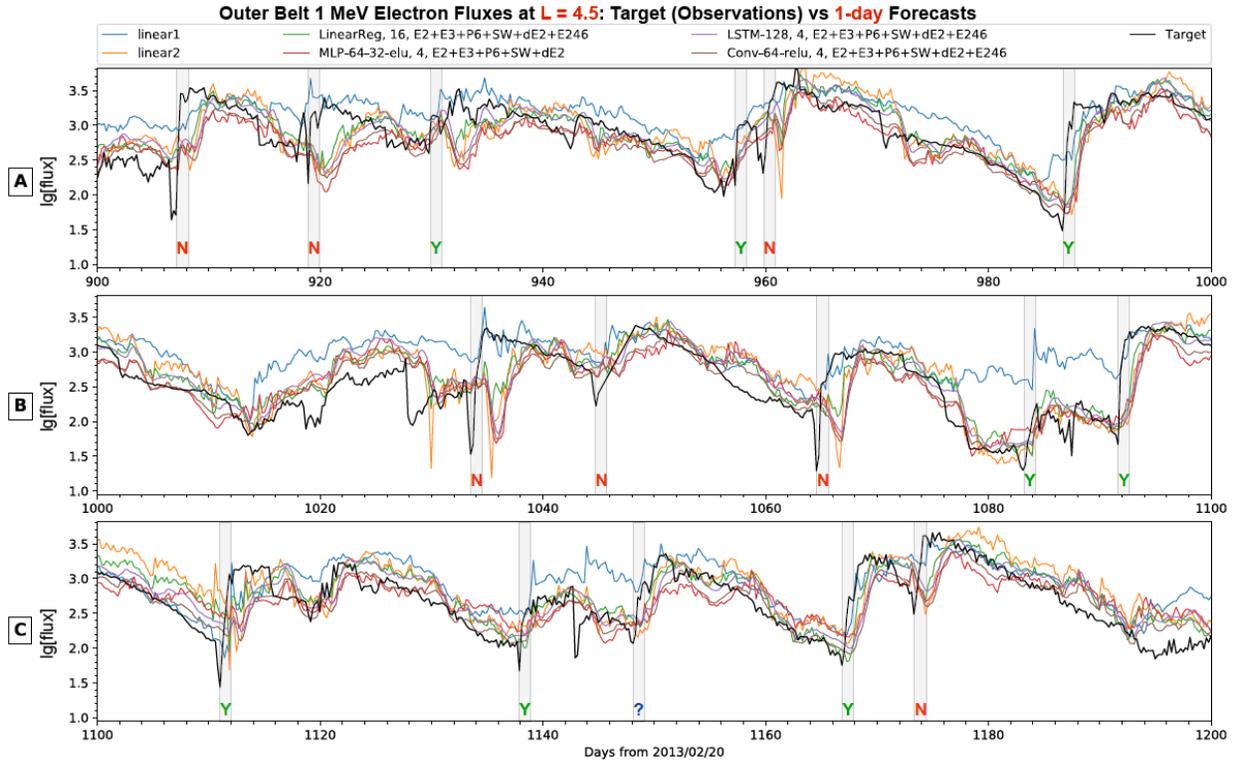

**Figure 10. One-day forecasts are compared to target fluxes at one single Lshell (L=4.5) over the validation and test period.** The time period is separated into three panels to show more details. Vertical gray boxes mark out 16 major MeV electron events—the left sides coincide the start of incoming MeV electron events and the width is 25 hr—and are also called prediction windows. A successful (failed, unclear) prediction of sudden MeV electron increment falls within (outside, on the edge) the prediction window and is marked with a green (red, blue) letter Y (N, ?).

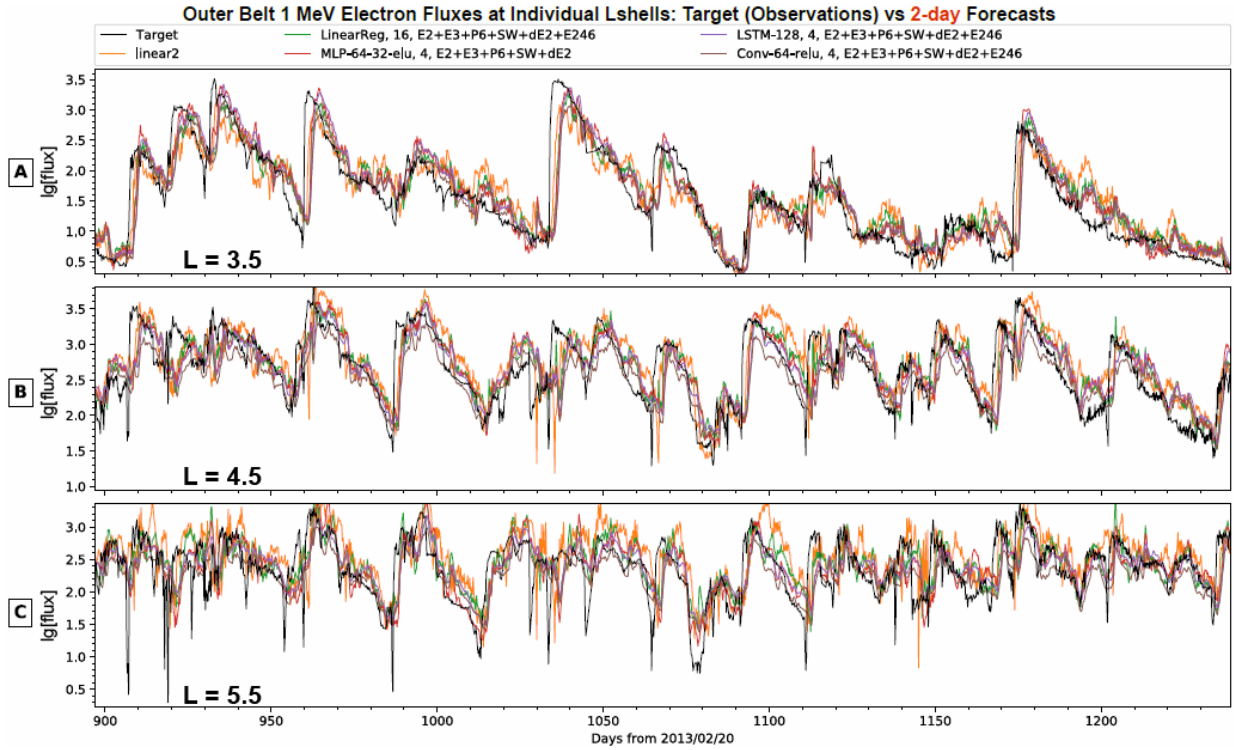

**Figure 11. Two-day forecasts are compared to target fluxes at three selected Lshells over the combined validation and test period.** Same format as Figure 9. Note here linear2 is for 1-day forecasts instead of 2-day.

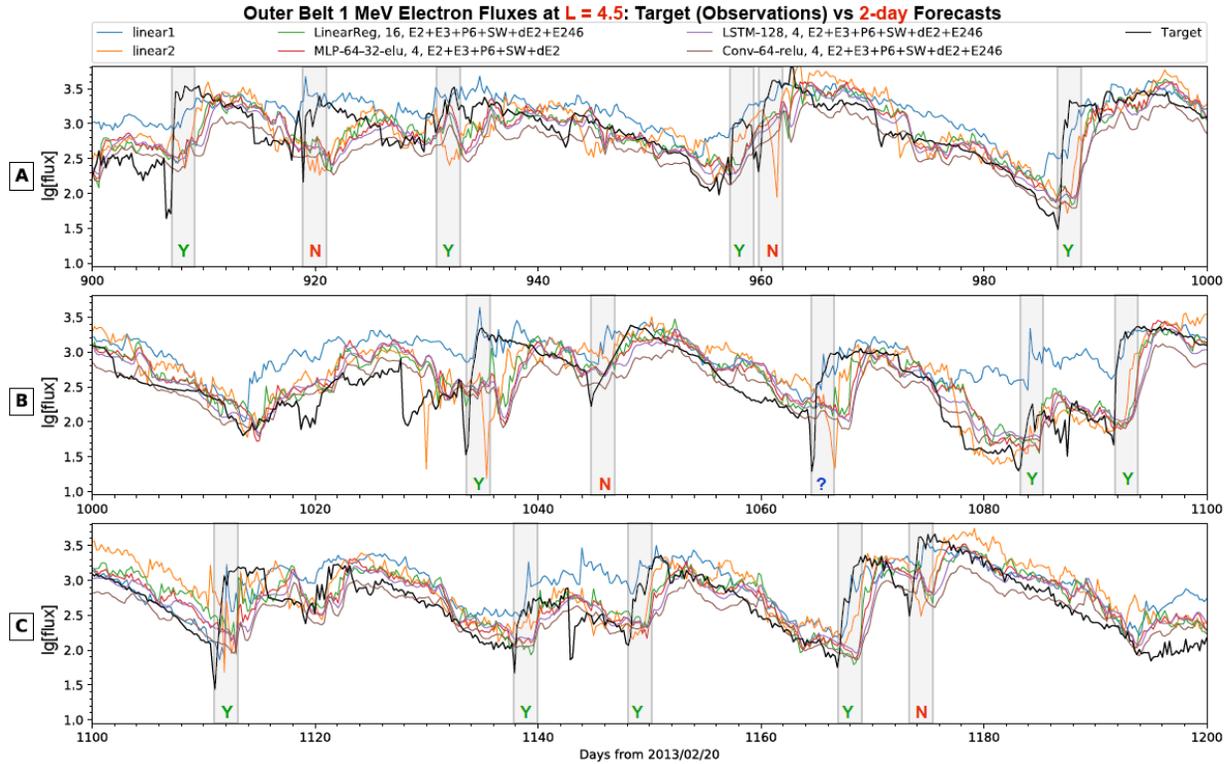

**Figure 12. Two-day forecasts are compared to target fluxes at one single Lshell (L=4.5) over the combined validation and test period.** Same format as Figure 10. The gray vertical boxes have a prediction window width of 50 hr. Note here linear1 and linear2 are for 1-day forecasts instead of 2-day.

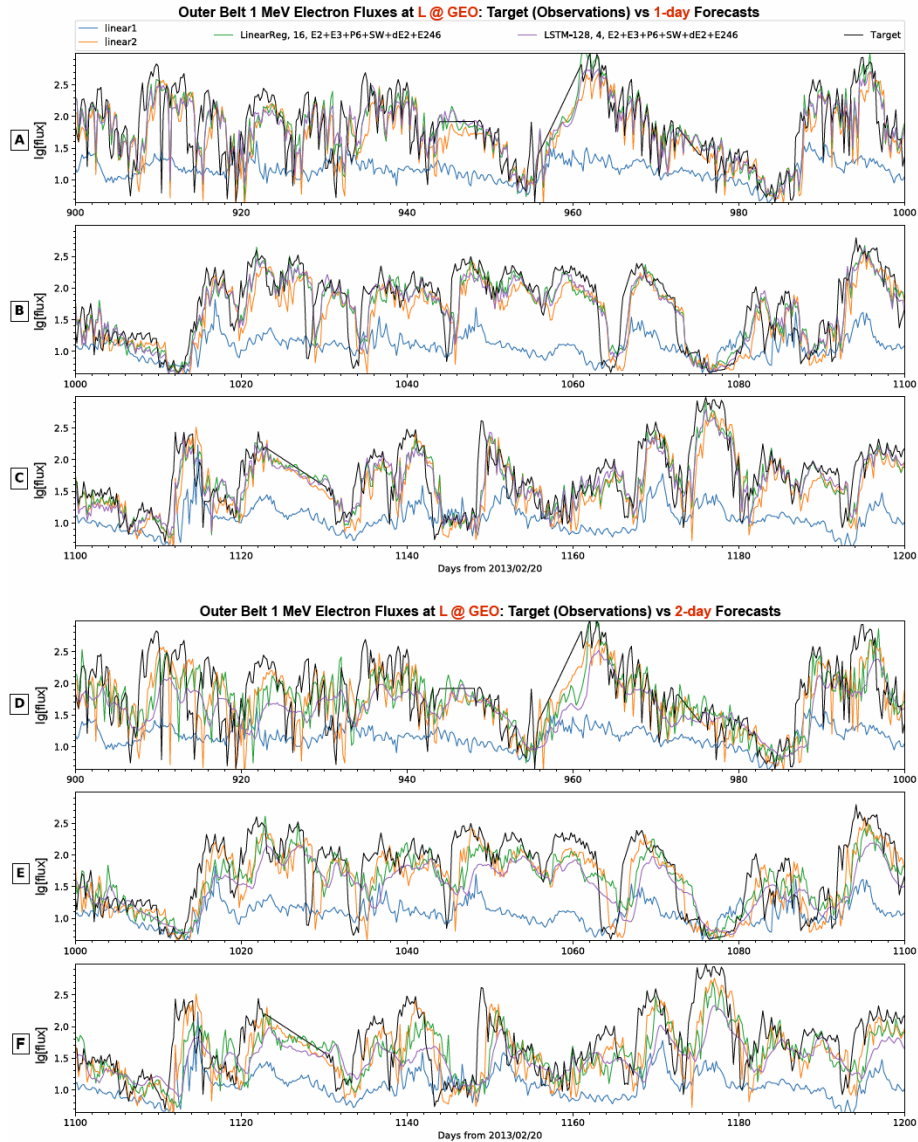

**Figure 13. One-day forecasts (A, B and C) and 2-day forecasts (D, E, and F) are compared to target fluxes at GEO over the validation and test period.** Same format as Figure 12. Here only results from LinearReg and LSTM models are shown for clearness. Note here linear1 and linear2 are all for 1-day forecasts.

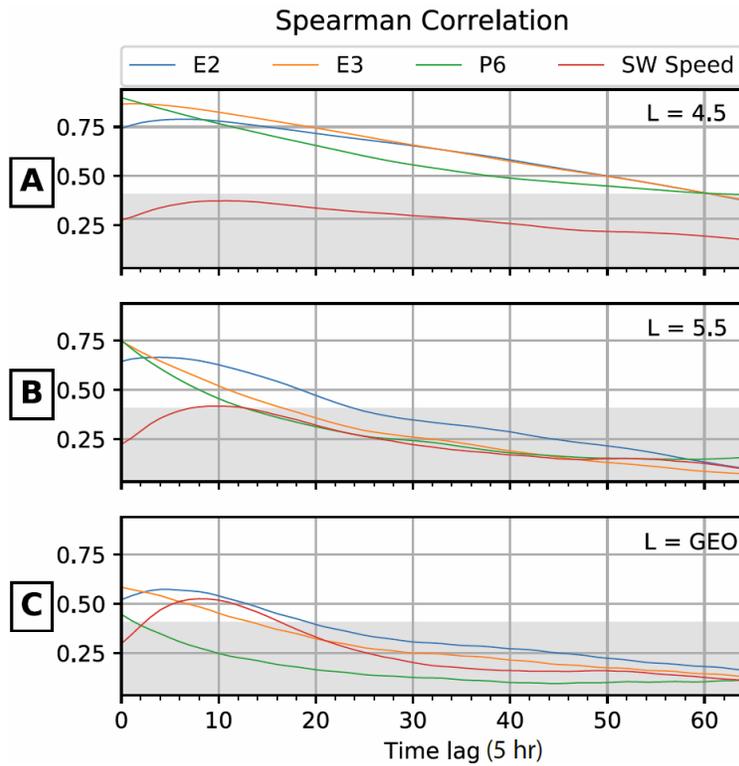

**Figure 14. Spearman correlation between target and input variables for multiple L-shells. A, B,** and **C** show, respectively, L=4.5, L=5.5, and at GEO the values of the Spearman correlation of E2, E3, P6, and the solar wind speed with the target 1 MeV electrons for different time lags. Each time lag corresponds to 5 hours. The top of each gray area corresponds to correlation value ~0.4.

**Tables**

**Table 1. Test input parameter combinations for 1-day (25 hr) forecasts.** Columns of PE values (averaged over all Lshells) are calculated for the training data set, validation data, test data, validation and test data combined together, and all data, respectively. The last column shows PE for validation and test combined at GEO only. The 10th model with the highest PE values is highlighted in red.

| model | window size | Input data | PE train | PE validation | PE test | PE val+test | PE all | PE GEO val+test |
|---|---|---|---|---|---|---|---|---|
| linear1 | 15 | E2/E3 | 0.679 | 0.602 | 0.695 | 0.668 | 0.691 | -1.732 |
| linear2 | 15 | E2+P6 | 0.869 | 0.753 | 0.816 | 0.797 | 0.854 | 0.352 |
| 01. LinearReg | 16 | SW | 0.569 | 0.289 | 0.634 | 0.518 | 0.574 | 0.557 |
| 02. LinearReg | 16 | E2 | 0.777 | 0.736 | 0.831 | 0.800 | 0.795 | 0.464 |
| 03. LinearReg | 16 | E3 | 0.822 | 0.769 | 0.846 | 0.822 | 0.831 | 0.390 |
| 04. LinearReg | 16 | P6 | 0.830 | 0.737 | 0.811 | 0.788 | 0.825 | 0.373 |
| 05. LinearReg | 16 | E2+E3 | 0.847 | 0.797 | 0.870 | 0.846 | 0.854 | 0.496 |
| 06. LinearReg | 16 | E2+P6 | 0.883 | 0.814 | 0.869 | 0.852 | 0.879 | 0.480 |
| 07. LinearReg | 16 | E2+SW | 0.782 | 0.747 | 0.839 | 0.809 | 0.801 | 0.574 |
| 08. LinearReg | 16 | E2+E3+P6 | 0.889 | 0.819 | 0.873 | 0.856 | 0.885 | 0.494 |
| 09. LinearReg | 16 | E2+E3+P6+SW | 0.896 | 0.830 | 0.874 | 0.861 | 0.890 | 0.584 |
| 10. LinearReg | 16 | E2+E3+P6+SW+dE2 | 0.897 | 0.830 | 0.875 | 0.861 | 0.891 | 0.587 |
| 11. LSTM-128 | 16 | SW | 0.503 | 0.052 | 0.551 | 0.382 | 0.490 | -1.471 |
| 12. LSTM-128 | 16 | E2 | 0.773 | 0.743 | 0.819 | 0.795 | 0.790 | 0.461 |
| 13. LSTM-128 | 16 | E3 | 0.805 | 0.753 | 0.832 | 0.807 | 0.815 | 0.394 |
| 14. LSTM-128 | 16 | P6 | 0.803 | 0.714 | 0.797 | 0.771 | 0.803 | 0.168 |
| 15. LSTM-128 | 16 | E2+E3 | 0.852 | 0.768 | 0.851 | 0.824 | 0.850 | 0.468 |
| 16. LSTM-128 | 16 | E2+P6 | 0.859 | 0.790 | 0.851 | 0.832 | 0.858 | 0.459 |
| 17. LSTM-128 | 16 | E2+SW | 0.793 | 0.777 | 0.826 | 0.812 | 0.809 | 0.548 |
| 18. LSTM-128 | 16 | E2+E3+P6 | 0.872 | 0.800 | 0.863 | 0.843 | 0.870 | 0.426 |
| 19. LSTM-128 | 16 | E2+E3+P6+SW | 0.886 | 0.829 | 0.866 | 0.855 | 0.882 | 0.536 |
| 20. LSTM-128 | 16 | E2+E3+P6+SW+dE2 | 0.884 | 0.826 | 0.865 | 0.853 | 0.880 | 0.564 |

**Table 2. Performance of models in four categories for 1-day (25 hr) forecasts.** Same format as Table 1. PE values for the top performer of each category are highlighted in red, also the top performers have their model index numbers marked with asterisk. E246 in the input list indicates E2 fluxes at L = 4.6.

| index | model | window size | input data | PE train | PE validation | PE test | PE val+test | PE all | PE GEO val+test |
|---|---|---|---|---|---|---|---|---|---|
| | linear1 | 15 | E2/E3 | 0.679 | 0.602 | 0.695 | 0.668 | 0.691 | -1.732 |
| | linear2 | 15 | E2+P6 | 0.869 | 0.753 | 0.816 | 0.797 | 0.854 | 0.352 |
| 1 | LinearReg | 4 | E2+E3+P6+SW | 0.890 | 0.827 | 0.871 | 0.858 | 0.886 | 0.568 |
| 2 | LinearReg | 16 | E2+E3+P6+SW | 0.896 | 0.830 | 0.874 | 0.861 | 0.890 | 0.584 |
| 3 | LinearReg | 4 | E2+E3+P6+SW+dE2 | 0.890 | 0.827 | 0.872 | 0.858 | 0.886 | 0.571 |
| 4 | LinearReg | 16 | E2+E3+P6+SW+dE2 | 0.897 | 0.830 | 0.875 | 0.861 | 0.891 | 0.587 |
| 5 | LinearReg | 4 | E2+E3+P6+SW+dE2+E246 | 0.902 | 0.838 | 0.884 | 0.869 | 0.897 | 0.571 |
| *6 | LinearReg | 16 | E2+E3+P6+SW+dE2+E246 | **0.906** | **0.838** | **0.887** | **0.872** | **0.900** | **0.587** |
| 7 | MLP-64-32-elu | 4 | E2+E3+P6+SW | 0.885 | 0.845 | 0.875 | 0.867 | 0.885 | 0.580 |
| 8 | MLP-64-32-elu | 16 | E2+E3+P6+SW | 0.901 | 0.830 | 0.875 | 0.861 | 0.894 | 0.575 |
| 9 | MLP-64-32-elu | 4 | E2+E3+P6+SW+dE2 | 0.874 | 0.841 | 0.871 | 0.863 | 0.877 | 0.502 |
| 10 | MLP-64-32-elu | 16 | E2+E3+P6+SW+dE2 | 0.888 | 0.834 | 0.871 | 0.860 | 0.885 | 0.363 |
| *11 | MLP-64-32-elu | 4 | E2+E3+P6+SW+dE2+E246 | **0.899** | **0.848** | **0.879** | **0.871** | **0.895** | 0.524 |
| 12 | MLP-64-32-elu | 16 | E2+E3+P6+SW+dE2+E246 | 0.904 | 0.833 | 0.878 | 0.864 | 0.897 | **0.590** |
| 13 | LSTM-128 | 4 | E2+E3+P6+SW | 0.877 | 0.837 | 0.874 | 0.863 | 0.879 | 0.448 |
| 14 | LSTM-128 | 16 | E2+E3+P6+SW | 0.884 | 0.822 | 0.862 | 0.850 | 0.879 | 0.530 |
| 15 | LSTM-128 | 4 | E2+E3+P6+SW+dE2 | 0.877 | 0.841 | 0.879 | 0.868 | 0.880 | 0.444 |
| 16 | LSTM-128 | 16 | E2+E3+P6+SW+dE2 | 0.886 | 0.830 | 0.870 | 0.859 | 0.883 | 0.558 |
| *17 | LSTM-128 | 4 | E2+E3+P6+SW+dE2+E246 | **0.893** | **0.846** | **0.886** | **0.874** | **0.892** | **0.589** |
| 18 | LSTM-128 | 16 | E2+E3+P6+SW+dE2+E246 | 0.899 | 0.830 | 0.872 | 0.859 | 0.892 | 0.536 |
| 19 | Conv-64-relu | 4 | E2+E3+P6+SW | 0.872 | 0.835 | 0.878 | 0.865 | 0.876 | 0.363 |
| 20 | Conv-64-32-relu | 16 | E2+E3+P6+SW | 0.799 | 0.696 | 0.791 | 0.761 | 0.797 | -0.121 |
| 21 | Conv-64-relu | 4 | E2+E3+P6+SW+dE2 | 0.872 | 0.838 | 0.884 | 0.870 | 0.877 | 0.352 |
| 22 | Conv-64-32-relu | 16 | E2+E3+P6+SW+dE2 | 0.787 | 0.664 | 0.776 | 0.740 | 0.783 | -0.112 |
| *23 | Conv-64-relu | 4 | E2+E3+P6+SW+dE2+E246 | **0.891** | **0.847** | **0.890** | **0.877** | **0.892** | **0.363** |
| 24 | Conv-64-32-relu | 16 | E2+E3+P6+SW+dE2+E246 | 0.796 | 0.677 | 0.770 | 0.741 | 0.789 | -0.308 |

**Table 3. Performance of models in four categories for 2-day (50 hr) forecasts.** Same format as Table 2. Note the PE values for linear 1 and linear2 are for 1-day forecasts instead of 2-day.

| index | model | window size | input data | PE train | PE validation | PE test | PE val+test | PE all | PE GEO val+test |
|---|---|---|---|---|---|---|---|---|---|
| | linear1 | 15 | E2/E3 | 0.679 | 0.600 | 0.695 | 0.667 | 0.691 | -1.723 |
| | linear2 | 15 | E2+P6 | 0.869 | 0.752 | 0.816 | 0.797 | 0.854 | 0.352 |
| 1 | LinearReg | 4 | E2+E3+P6+SW | 0.853 | 0.772 | 0.837 | 0.817 | 0.849 | 0.269 |
| 2 | LinearReg | 16 | E2+E3+P6+SW | 0.858 | 0.774 | 0.841 | 0.820 | 0.854 | 0.323 |
| 3 | LinearReg | 4 | E2+E3+P6+SW+dE2 | 0.853 | 0.772 | 0.838 | 0.818 | 0.850 | 0.282 |
| 4 | LinearReg | 16 | E2+E3+P6+SW+dE2 | 0.860 | 0.773 | 0.842 | 0.820 | 0.855 | 0.333 |
| 5 | LinearReg | 4 | E2+E3+P6+SW+dE2+E246 | 0.865 | 0.779 | 0.844 | 0.824 | 0.859 | 0.282 |
| *6 | LinearReg | 16 | E2+E3+P6+SW+dE2+E246 | **0.871** | **0.778** | **0.849** | **0.827** | **0.864** | **0.333** |
| 7 | MLP-64-32-elu | 4 | E2+E3+P6+SW | 0.846 | 0.775 | 0.838 | 0.819 | 0.845 | 0.102 |
| 8 | MLP-64-32-elu | 16 | E2+E3+P6+SW | 0.859 | 0.765 | 0.829 | 0.809 | 0.851 | 0.285 |
| *9 | MLP-64-32-elu | 4 | E2+E3+P6+SW+dE2 | **0.857** | **0.780** | **0.842** | **0.823** | **0.854** | 0.200 |
| 10 | MLP-64-32-elu | 16 | E2+E3+P6+SW+dE2 | 0.863 | 0.772 | 0.837 | 0.817 | 0.855 | 0.326 |
| 11 | MLP-64-32-elu | 4 | E2+E3+P6+SW+dE2+E246 | 0.860 | 0.776 | 0.835 | 0.817 | 0.854 | -0.037 |
| 12 | MLP-64-32-elu | 16 | E2+E3+P6+SW+dE2+E246 | 0.859 | 0.772 | 0.832 | 0.814 | 0.852 | **0.353** |
| 13 | LSTM-128 | 4 | E2+E3+P6+SW | 0.835 | 0.760 | 0.833 | 0.810 | 0.835 | 0.073 |
| 14 | LSTM-128 | 16 | E2+E3+P6+SW | 0.847 | 0.757 | 0.825 | 0.804 | 0.842 | 0.208 |
| 15 | LSTM-128 | 4 | E2+E3+P6+SW+dE2 | 0.840 | 0.767 | 0.837 | 0.815 | 0.840 | 0.119 |
| 16 | LSTM-128 | 16 | E2+E3+P6+SW+dE2 | 0.851 | 0.762 | 0.831 | 0.810 | 0.846 | 0.307 |
| *17 | LSTM-128 | 4 | E2+E3+P6+SW+dE2+E246 | **0.853** | **0.770** | **0.839** | **0.818** | **0.849** | 0.104 |
| 18 | LSTM-128 | 16 | E2+E3+P6+SW+dE2+E246 | 0.865 | 0.765 | 0.830 | 0.811 | 0.855 | **0.353** |
| 19 | Conv-64-relu | 4 | E2+E3+P6+SW | 0.841 | 0.771 | 0.840 | 0.819 | 0.842 | 0.080 |
| 20 | Conv-64-32-relu | 16 | E2+E3+P6+SW | 0.765 | 0.607 | 0.742 | 0.699 | 0.756 | -0.498 |
| 21 | Conv-64-relu | 4 | E2+E3+P6+SW+dE2 | 0.840 | 0.761 | 0.839 | 0.814 | 0.840 | 0.056 |
| 22 | Conv-64-32-relu | 16 | E2+E3+P6+SW+dE2 | 0.769 | 0.596 | 0.731 | 0.688 | 0.755 | -0.512 |
| *23 | Conv-64-relu | 4 | E2+E3+P6+SW+dE2+E246 | **0.846** | **0.767** | **0.844** | **0.820** | **0.845** | -0.073 |
| 24 | Conv-64-32-relu | 16 | E2+E3+P6+SW+dE2+E246 | 0.771 | 0.583 | 0.715 | 0.673 | 0.753 | -0.636 |